\documentclass[aps,pra,twocolumn,superscriptaddress,showpacs,aps]{revtex4}
\usepackage{amsmath,epsfig,amsfonts,graphicx}
\usepackage{subfigure}
\newtheorem{theorem}{Theorem}[section]
\begin{document}

\title{Finite-temperature dynamics of matter-wave dark solitons in linear and periodic potentials: 
an example of an anti-damped Josephson junction}
\author{Y. Shen}
\affiliation{Department of Mathematics and Statistics, University of Massachusetts,
Amherst, MA 01003-4515, USA}
\author{P. G. Kevrekidis}
\affiliation{Department of Mathematics and Statistics, University of Massachusetts,
Amherst, MA 01003-4515, USA}
\author{N. Whitaker}
\affiliation{Department of Mathematics and Statistics, University of Massachusetts,
Amherst, MA 01003-4515, USA}
\author{N. I. Karachalios}
\affiliation{Department of Mathematics, University of the Aegean, Karlovassi, 83200 Samos, Greece}
\author{D. J. Frantzeskakis}
\affiliation{Department of Physics, University of Athens, Panepistimiopolis, Zografos, Athens 15784, Greece}

\begin{abstract}
We
study matter-wave dark solitons in atomic Bose-Einstein
condensates
at finite temperatures, under the effect of linear and periodic potentials.
Our model, namely a dissipative Gross-Pitaevskii equation, is treated analytically by means of
dark soliton perturbation theory, which results in a Newtonian equation of motion for the dark soliton center.
This reduced model, which incorporates
an effective washboard potential and an anti-damping term, constitutes an example of an anti-damped Josephson junction.
We present a qualitative (local and global) analysis of the equation of motion. For sufficiently small wavenumbers of the periodic potential and weak linear potentials, the results are found to be in good agreement with pertinent ones obtained via a Bogoliubov-de Gennes analysis and direct numerical simulations.

%

\end{abstract}


\maketitle


\section{INTRODUCTION}

During the last years, matter-wave dark solitons have been the focus of many research efforts in
the physics of ultracold atoms, due to their ubiquitous presence in the formation and nonlinear dynamics of
atomic Bose-Einstein condensates (BECs) \cite{book,review}.
%
%
Importantly, these efforts have been driven by considerable experimental developments along this direction. In particular,
experimental investigations on matter-wave dark solitons started over one decade ago~\cite{han1,nist,dutton,han2};
however, since the soliton lifetime was rather limited
(either due to the instability of dark solitons in higher-dimensional
setups, or due to the detrimental effect of strong thermal fluctuations),
these early experiments were not able to characterize many of the soliton properties.
Nevertheless, there has been a new generation of more controllable experiments
\cite{engels,hamburg,hambcol,kip,andreas},
which have been carried out at
extremely low temperatures and --at least in some cases-- in more quasi one-dimensional (1D) setups,
so as to enable the definitive observation of robust moving, oscillating and
potentially interacting dark solitons,
in very good agreement with the corresponding theoretical predictions.
Furthermore, dark solitons were also observed recently in multi-component BECs, where
they were coupled either with dark or bright solitons, thus
forming ``dark-bright''~\cite{engels1,engels2,engels3,engels4}
or ``dark-dark''~\cite{engels3,beating}
soliton states, respectively.
%

In the same context, and since BEC experiments are obviously performed in the presence of
thermal fluctuations, the study of matter-wave dark solitons in finite-temperature BECs is
a quite relevant and interesting problem. Pertinent theoretical studies have been performed in various settings and using different approaches (see, e.g., Refs.~\cite{fms,npppd,npp1,Cockburn,kamenev,cock2,bradley,us2,db2}).
A relevant -- and analytically tractable -- mean-field model that has recently gained attention
\cite{npppd,Cockburn,cock2,bradley,us2,db2} (also in the context of vortices \cite{stephan}) is
the so-called {\it dissipative Gross-Pitaevskii equation} (DGPE). This model, which was first introduced
phenomenologically \cite{lp} and later was justified from a microscopic perspective \cite{npprev},
can describe accurately finite-temperature-induced soliton decay: results stemming from a perturbative
study of soliton dynamics in the framework of DGPE, compares favorably to ones obtained by
the more accurate {\it stochastic Gross-Pitaevskii} model \cite{Cockburn,cock2,bradley} (the latter, is a
grand canonical theory of thermal BECs, containing damping and noise terms which describe
interactions of low-energy atoms with a high-energy thermal reservoir \cite{npprev}).
%
%
%

Our aim in the present work is in a sense two-fold. On the one hand,
we aim to study matter-wave dark solitons in thermal BECs
in an experimentally relevant and realizable setting. On the other hand, we are
aiming to reverse engineer an interesting dynamical system that has been studied in the context
of damping (and external constant driving) in some detail in the context
of Josephson junctions; see relevant details in Ch.~2 of Ref.~\cite{dau} and
Ch.~8 of Ref.~\cite{strogatz} for physical and mathematical aspects, respectively.
The present setting of BECs enables the possibility of
analyzing this dynamical system in the presence of an anti-damping mechanism
induced by the coupling of the dark soliton to thermal fluctuations.

In particular, we focus on the examination of matter-wave dark soliton dynamics in the presence of
the following effects:
%
%
(i) coupling to thermal excitations, described in the framework
of the DGPE model discussed above;
%
%
(ii) a (weak) periodic, so-called optical lattice (OL)~\cite{markus}, potential \cite{note};
%
%
(iii) an (also weak) electric or gravitational field, inducing a linear external potential
-- see, e.g., Ref.~\cite{bpa} for an experimental realization.
%
%

The steps of our analysis, and the structure of our presentation, are as follows.
First, starting
from a
DGPE which encompasses the
above potentials (linear and periodic), we employ
dark soliton perturbation theory~\cite{review,Kivshar,gt,Theocharis}, to derive an equation of motion
for the ensuing dynamics of the soliton center position
(section II). Motivated by its direct
analogy with the equation describing Josephson junction dynamics
(which has exciting features such multi-stability and hysteresis),
we will offer a systematic local and global analysis (fixed points and phase plane structure)
of the pertinent dynamical system.
Then (in section III), we will present results of systematic numerical simulations,
comparing our analytical results with ones obtained from direct numerical integration of the DGPE.
Finally (in section IV), we will summarize our results
and present our conclusions, as well as some interesting possibilities
for future extensions of this work.

\section{THEORETICAL APPROACH}
\label{THA}

\subsection{Perturbation theory for dark solitons.}

The dynamics of a quasi-1D condensate (elongated along the $x$-direction) under the presence of thermal
fluctuations, can be described by the following dimensionless DGPE \cite{Cockburn,cock2}:
%
%
\begin{equation}
(i-\gamma)u_t+\frac{1}{2}u_{xx}-|u|^2u+\mu u-V(x)u = 0.
\label{GP}
\end{equation}
Here, subscripts denote partial derivatives, $u(x,t)$ is the condensate wavefunction, $\mu$ is the chemical potential,
and the term $-\gamma u_t$ accounts for finite-temperature effects (note that the parameter $\gamma$ scales with temperature $T$ according to a power law, $\gamma \propto T^{\alpha}$,
with $1<\alpha < 4$) \cite{Cockburn,cock2,bradley}.
%
%
Finally, the external potential $V(x)$ is assumed to of the form:
\begin{equation}
V(x) = -Ex+ V_0 \cos(2kx),
\label{extpot}
\end{equation}
where, in the right-hand side, the first term
%
%
represents an electric or gravitational field of strength $E$, and the second term
%
%
describes an optical lattice potential of strength $V_0$ and wavenumber $k$ \cite{bpa}.
%

Without the external potential and the thermal term, i.e., when $V(x)=0,\ \gamma=0$, Eq.~(\ref{GP})
possesses a dark soliton solution on top of a constant background of density $\mu$; this solution is
%
%
of the form
\begin{equation}
u(x,t) = \sqrt{\mu}[\cos\phi \tanh(X) + i \sin\phi ],
\label{ds_unp}
\end{equation}
where $X= \sqrt{\mu}\cos\phi[x-x_0(t)]$ and $x_0(t)=\sqrt{\mu}(\sin\phi) t$ is the soliton center.
The amplitude and velocity of the soliton are respectively given by
$\sqrt{\mu}\cos\phi$ and $\sqrt{\mu} \sin\phi$, while $\phi$ (with $|\phi| < \pi/2$)
is the soliton phase angle (so-called ``black'' and ``grey'' solitons correspond to $\phi=0$ and $\phi \ne 0$, respectively).

Following Refs.~\cite{Cockburn,cock2}, we now seek a solution of Eq. (\ref{GP}) in the form
$u(x,t) =u_b(x)\exp[-i\theta(t)]v(x,t)$, where $u_b(x)$ and $\theta(t)$
denote the background amplitude and phase, respectively, while the unknown complex function
$v(x,t)$ denotes a dark soliton. We assume that the condensate dynamics involves a fast
relaxation scale to the ground state, which can be approximated in the
framework of the Thomas-Fermi (TF) approximation, as follows:
$u_b \approx \sqrt{{\rm max}\left\{\mu-V(x), 0\right\}}$. Then, the evolution
of the dark soliton on top of this ground state, is described by the following
perturbed nonlinear Schr\"{o}dinger (NLS) equation:
%
%
%
%
\begin{equation}
iv_t+\frac{1}{2}v_{xx}-(|v|^2-1)v =P[v],
\label{perturb}
\end{equation}
where we have used the scale transformations
$t\rightarrow\mu t$, $x\rightarrow\sqrt{\mu} x$, and $V \rightarrow V/\mu$,
and the functional perturbation $P[v]$ is given by:
%
%
\begin{equation}
P[v] = (1-|v|^2)vV+\frac{1}{2}V_x v_x+\gamma v_t.
\end{equation}
Below we will treat $P[v]$ as a small perturbation (at least in the spatial region
where the dark soliton dynamics takes place), assuming that the thermal fluctuations
(governed by the parameter $\gamma$) and the strengths $V_0$ and $E$ of the linear and OL
potentials are weak, namely,
%
%
$\gamma \ll 1$, $V_0 \ll 1$ and $E \ll 1$.
Under these
assumptions,
we can then apply to Eq.~(\ref{perturb}) the perturbation theory for dark solitons,
originally developed in Ref.~\cite{Kivshar}
and then adapted to dark solitons in BECs in Ref.~\cite{gt} (see also Ref.~\cite{Theocharis}, for an
application to BECs confined in OLs, and Ref.~\cite{review} for a review).

The fundamental premise underlying the dark soliton perturbation
theory is that, to the leading order of approximation, the soliton evolves adiabatically;
in other words, it preserves its form of Eq.~(\ref{ds_unp}),
yet its parameters, namely the phase angle $\phi$ and the position of the center $x_0$ become
slowly-varying functions of time.
More specifically, the solution of Eq.~(\ref{perturb}) is taken to be of the form
$v = \cos\varphi(t) \tanh[\cos\varphi(t)(x-x_0(t))]+i\sin\varphi(t)$, where
$x_0(t)=\int_0^t\sin\varphi(t){\rm d}t$ and $\varphi(t)$ denote the
time-dependent center and phase angle of the dark soliton, respectively.
%
%
These functions are found to evolve according to the following equations
%
%
\cite{review,Kivshar,gt,Theocharis}:
\begin{eqnarray}
\dot{\varphi}&=& \frac{1}{2\cos^2{\varphi}\sin{\varphi}}
{\rm Re}\Big\{\int_{-\infty}^{+\infty} P[v]v^*_t {\rm d}x\Big\},
\label{perturbeqn1} \\
\dot{x}_0&=&\sin\varphi(t),
\label{perturbeqn2}
\end{eqnarray}
where overdots denote time derivatives and asterisk denotes complex conjugate.
Evaluating the integral in Eq.~(\ref{perturbeqn1}), and
considering the case of almost black solitons with sufficiently small soliton phase angles,
we neglect terms of order $\mathcal{O}(d\varphi/dt)$. Then, we can derive from
Eqs.~(\ref{perturbeqn1})-(\ref{perturbeqn2}) an equation for the soliton center;
this equation, when expressed in the original variables, has the following form:
%
%
%
%
%
\begin{eqnarray}
\ddot{x}_0&=& \frac{2}{3} \gamma \mu
\dot{x}_0+ \frac{1}{2}  E
+ A_k k V_0 \sin(2 k x_0),
\label{eqn_center_original}
\end{eqnarray}
where
\begin{equation}
A_k \equiv \frac{\pi k}{\sqrt{\mu}} \left(1 + \frac{k^2}{\mu}\right) {\rm csch}\left(\frac{\pi k}{\sqrt{\mu}}\right),
\label{Ak}
\end{equation}
and $A_k\approx1$ for small $k$. Note that Eq.~(\ref{eqn_center_original}) has the form of a Newtonian equation of motion
of a classical unit-mass particle, in the presence of (anti)damping, of strength $\propto (2/3)\gamma \mu$, and in the
presence of the effective ``washboard'' potential:
\begin{eqnarray}
\label{potdark0}
U(x)=\frac{1}{2}\left(
A_k V_0 \cos(2k x)
-E x\right).
\end{eqnarray}
Alternatively, for the purposes of our analytical considerations,
Eq.~(\ref{eqn_center_original}) can be expressed as the following dynamical system:
%
%
\begin{eqnarray}
\dot{x} &=&
y,
\nonumber \\
\dot{y} &=&
\frac{2}{3}\gamma\mu y+\frac{1}{2}E+A_k kV_0 \sin(2kx).
\label{eqn_center_dark_system}
\end{eqnarray}

It is quite remarkable that
this dynamical system draws a direct analogy between
the dynamics
of a dark soliton in a thermal BEC under the combined
effect of gravity and an optical lattice, and
that of a {\it Josephson junction}. In the latter, the
dynamical variable is the relative phase between the two superconducting
elements, whose time derivative is associated with the voltage in the
junction. The bias current is analogous to our linear field here, while
the Josephson current is the one that provides the periodic potential.
Lastly, the lossy resistor part in the original Josephson junction setting (i.e., $\gamma<0$)
is replaced here by the {\it anti-damping} effect (i.e., $\gamma>0$) induced by thermal fluctuations.
It is also quite noteworthy that the original experimental setting
of Ref.~\cite{bpa} (see also Ref.~\cite{stein}), which is the closest
to our present proposal, was precisely
designed to examine macroscopic quantum interference between the atomic
packets in adjacent wells of the lattice in close analogy to the so-called
alternating-current (ac) Josephson effect. Here, we take this analogy one step further by
parallelizing the motion of a dark soliton through this potential environment
to the evolution of the phase of the Josephson junction.

%
%

\subsection{Qualitative analysis of the dynamics}
\label{QA}

Let us now analyze in detail the dynamics of the system of Eqs.~(\ref{eqn_center_dark_system}).
First, we note that due to the periodicity of the nonlinearity in the system~(\ref{eqn_center_dark_system}), we will restrict the analysis to the interval $0\le x \le \pi/k$ (one period of the periodic function within effective potential).
Second, it is relevant to observe that two different possibilities occur:
%
%
%
the {\em existence} and {\em non-existence} of fixed points
of Eqs.~(\ref{eqn_center_dark_system}). Obviously, fixed points exist for
$E/2A_k kV_0\leq 1$, i.e., for
%
%
$E/2A_k kV_0<1$ (existence of two fixed points) or
%
%
$E/2A_k k V_0=1$ (existence of one fixed point). First, we will provide the local analysis for the case $E/2A_k kV_0<1$. This case, as well as the one with $E/2A_k k V_0=1$, and the scenario where no fixed points exist, will be investigated in the next section devoted to the global analysis of the dynamics.

%
%
%
%

\subsubsection{Local analysis}

In the case $E/2A_k kV_0<1$, there exist {\it two} fixed points:
$\bar{\mathbf{x}}_{-}=(\bar{x}_{-}, 0)$ and $\bar{\mathbf{x}}_{+}=(\bar{x}_{+}, 0)$.
%
%
The $x$-coordinates of the fixed points  $\bar{\mathbf{x}}_{\pm}$
satisfy the transcendental equation
\begin{equation}
\sin(2k\bar{x}_{\pm})=-\frac{E}{2A_k kV_0},
\label{tran}
\end{equation}
and correspond to the local minimum and the local maximum of the potential energy $U(x)$ respectively. To study the stability of the two fixed points, we first
consider the linearization of the system (\ref{eqn_center_dark_system}) around them,
whereby the Jacobian matrix reads:
\begin{eqnarray}
J\big|_{\bar{\mathbf{x}}_{\pm}}
= \left( \begin{array}{ccc}
  0 & 1\\
  \pm \alpha & \frac{2}{3}\gamma\mu
  \end{array}\right),
\label{jac}
\end{eqnarray}
where $\alpha= k(4A_k^2k^2V_0^2-E^2)^{1/2}$.
The fixed point $\bar{\mathbf{x}}_+$ [where its $x$-coordinate, $\bar{x}_{+}$, is the maximum of the potential energy $U(x)$] has eigenvalues:
\begin{equation}
\lambda_{1,2}^{(+)}(\gamma) = \frac{1}{3}\gamma\mu\pm \sqrt{\frac{1}{9}\gamma^2\mu^2+\alpha}.
\label{eig1}
\end{equation}
%
%
%
Since $\alpha>0$, it is clear that $\lambda_{1}^{(+)}>0$ and $\lambda_{2}^{(+)}<0$ and, hence,
$\bar{\mathbf{x}}_+$ is always unstable: in fact, it is a {\it saddle point},
independently of whether $\gamma=0$
(corresponding to the Hamiltonian variant of the model) or $\gamma>0$
(corresponding to the finite-temperature case).
%

On the other hand, the fixed point $\bar{\mathbf{x}}_-$ [where its $x$-coordinate, $\bar{x}_{+}$, is the minimum of the potential energy $U(x)$] has eigenvalues:
\begin{equation}
\lambda_{1,2}^{(-)}(\gamma) = \frac{1}{3}\gamma\mu\pm \sqrt{\frac{1}{9}\gamma^2\mu^2-\alpha}.
\label{eig2}
\end{equation}
In the Hamiltonian case $\gamma=0$ (where the Hamiltonian energy $\mathcal{H}(x,y) \equiv \frac{1}{2}y^2+U(x)$
is conserved), the above eigenvalues are purely imaginary,
i.e., $\lambda_{1,2}^{(-)} = \pm i \alpha$, and thus $\bar{\mathbf{x}}_{-}$ is a
%
%
%
{\it stable center}; in this case, the dark soliton center performs oscillatory motions around the minimum of the effective potential.
However, in the finite-temperature case of $\gamma>0$ (where the Hamiltonian energy
$\mathcal{H}(x,y)$ is not conserved), we need to separate two cases.
If $\frac{1}{9}\gamma^2\mu^2<\alpha$ (i.e., for sufficiently
low temperatures), the eigenvalues are complex with positive real part and, thus,
the fixed point $\bar{\mathbf{x}}_{-}$ is an {\it unstable spiral}. Physically, this means that
the dark soliton center will perform a motion with an oscillatory growing amplitude
around the minimum of the effective potential,
with a growth rate
$\frac{1}{3}\gamma\mu$ and an oscillation frequency:
\begin{equation}
\omega_{\rm eff} = \sqrt{\alpha-\frac{1}{9}\gamma^2\mu^2}.
\label{efom}
\end{equation}
%
%
On the other hand, if $\gamma$ is sufficiently large (such that
$\frac{1}{9}\gamma^2\mu^2>\alpha$), the fixed point $\bar{\mathbf{x}}_{-}$ becomes an {\it unstable node}.
Physically, this means that the dark soliton center will perform a motion with an exponentially growing
amplitude. Note that, in both cases, the soliton will eventually decay at the rims of the condensate.

At this point, we should mention that the local analysis highlights
the emergence of, at least, two different bifurcations. First, the
transition from $\gamma=0$ to finite $\gamma>0$ is a bifurcation whereby the minimum of the effective potential
changes stability: it becomes from stable center an unstable spiral, with the eigenvalues becoming, from
purely imaginary,
genuinely
complex with positive real part.
Second, if $\gamma$ is further increased so as to exceed the
critical value
\begin{equation}
\gamma_{\rm cr} = \frac{3\sqrt{\alpha}}{\mu},
\label{gcr}
\end{equation}
the character of the instability changes via another bifurcation:
the fixed point, from unstable spiral becomes an unstable node, with the eigenvalues turning, from genuinely complex, to real positive
(via a collision on the real axis). However, the global analysis
that will be discussed in the next section will reveal,
in the regime $0<\gamma<\gamma_{\rm cr}$,
a third type of bifurcation: a saddle connection breaking, which is relative to the behavior
of the unstable manifolds of the saddle point $\bar{\mathbf{x}}_{-}$.
%

To conclude this section, we will now consider the case $E/2A_k k V_0=1$, as well as the
scenario where no fixed points exist. In the first case, there exists solely one fixed point, say $(\bar{x},0)$, which
occurs from the collision between $(\bar{x}_-,0)$ and $(\bar{x}_+,0)$ when $\alpha=0$. Then, the
Jacobian of Eq.~(\ref{jac}) has eigenvalues $\lambda_1=0$ and $\lambda_2 = (2/3)\gamma \mu>0$. In this case,
the linear stability analysis is insufficient to
provide any information about the stability of $\bar{x}$.
Both the critical case and that of the absence of fixed points
will be examined  by
means of global analysis of the system -- see below.

\subsubsection{Global analysis}

Let us now study the global phase portrait of the system (\ref{eqn_center_dark_system}).
We will use two fundamental theorems
for flows in two-dimensional phase spaces, namely
the {\it Hopf bifurcation theorem} (see, e.g., Theorem 3.4.2
in Ref.~\cite{JGPH}) and the
{\it Poincar\'{e}-Bendixson theorem} (see, e.g., Theorem 1.1.19
in Ref.~\cite{W90}),
combined with a {\it nullcline analysis}.
The latter, is an analysis of the direction of the flow around the curves consisting of the points on which the vector field is parallel ($\dot{y}=0$), or
perpendicular ($\dot{x}=0$) to the horizontal axis $y=0$ \cite{HSD}.
%

First we consider the case $E/2A_k k V_0<1$, where the two fixed points $\bar{\mathbf{x}}_{-}=(\bar{x}_{-},0)$ and $\bar{\mathbf{x}}_{+}=(\bar{x}_{+},0)$ exist.
Since the potential energy $U(x)$ [cf. Eq.~(\ref{potdark0})] is decreasing,
the maxima (minima) $\bar{x}_{+}~\mathrm{mod}(\pi/k)$ [$\bar{x}_{-}~\mathrm{mod}(\pi/k)$]
are in different and decreasing energy levels, with every minimum lying between two maxima.
Therefore, in the Hamiltonian case, $\gamma=0$, the conditions for the existence of homoclinic orbits
(cf. Lemma 14.3
in Ref.~\cite{H91}) are satisfied;
thus, at the energy levels of the maxima $\bar{x}_{+}~\mathrm{mod}(\pi/k)$
of $U(x)$, and in the direction of increasing energy, correspond to homoclinic orbits on the saddles $\bar{\mathbf{x}}_{+}~\mathrm{mod}(\pi/k)$, surrounding the centers $\bar{\mathbf{x}}_{-}~\mathrm{mod}(\pi/k)$.
Let us denote by $W^s(\bar{\mathbf{x}}_{+})$ and $W^u(\bar{\mathbf{x}}_{+})$, the stable and the unstable manifold of $\bar{\mathbf{x}}_{+}$, respectively. We may perform a nullcline analysis in the rectangular region
$B_0=\left\{0\leq x\leq\frac{\pi}{k},\;Y_1\leq y\leq Y_2\right\}$,
where $Y_1<0$ and $y=Y_2>0$ are arbitrary horizontal lines, below and above the homoclinic orbit of $\bar{\mathbf{x}}_{+}$, respectively.
%
%
This analysis shows that the non-homoclinic branch of $W^u(\bar{\mathbf{x}}_{+})$
escapes from $B_0$ and becomes unbounded. On the other hand, in the Hamiltonian case $\gamma=0$, the orbits are symmetric with respect to the
horizontal axis $y=0$. The symmetric orbit of the unbounded branch of $W^u(\bar{\mathbf{x}}_{+})$ is the non-homoclinic branch of $W^s(\bar{\mathbf{x}}_{+})$, which enters in the region $B_0$, by crossing $y=Y_2$ from below.

The transition from $\gamma=0$ to finite $0<\gamma<\gamma_{\mathrm{cr}}$, can be analyzed with the help of the Hopf bifurcation theorem and the Poincar\'{e}-Bendixson theorem. The former, establishes
the non-existence of periodic orbits bifurcating from the unstable spiral $\bar{\mathbf{x}}_{-}$.
This is due to the fact that, while the two conditions of the theorem concerning the eigenvalues are satisfied,
the third one concerning the degeneracy condition is not (see, e.g., Ref.~\cite{JGPH}).
%
%
Besides, the nullcline analysis
in region $B_0$ shows that, apart from the stable manifold $W^s(\bar{\mathbf{x}}_{+})$, all orbits
%
escape form the strip $Y_1\leq y\leq Y_2$, and
there is not any other closed orbit in the strip.

Since a periodic orbit does not exist,
one may apply the Poincar\'{e}-Bendixson theorem on the
invariant manifold $W^s(\bar{\mathbf{x}}_{+})$. This way, we can
establish that each branch lying in the half-plane $y>0$ joins the unstable spiral $\bar{\mathbf{x}}_{+}$.
The situation is more intriguing for the branch of $W^s(\bar{\mathbf{x}}_{+})$ lying in the half-plane $y<0$,
since two possibilities exist regarding its connection
to a fixed point.
These possibilities
depend on the size of $0<\gamma<\gamma_{\rm cr} = 3\sqrt{\alpha}/\mu$,
and on the fact, that the transition from $\gamma=0$ to finite $\gamma>0$ is a homoclinic bifurcation for the saddle $\bar{\mathbf{x}}_{+}$, where each homoclinic orbit for $\gamma=0$ breaks.

When $\gamma<\gamma_{\rm cr}$ is sufficiently small, a nullcline analysis for $\gamma \rightarrow 0$,
shows that the gradient of the branch $W^s(\bar{\mathbf{x}}_{+})$ lying in the half-plane $y<0$
is close to
 that of its analogue for $\gamma=0$; furthermore, for $\gamma \rightarrow 0$,
the spiraling growth is weak. Therefore, since the branch of $W^s(\bar{\mathbf{x}}_{+})$
lying in the half-plane $y>0$ is connected
to the spiral point $\bar{\mathbf{x}}_{-}$, the other branch of $W^s(\bar{\mathbf{x}}_{+})$ lying in the half-plane $y<0$,
is the one that connects to the saddle point $\bar{\mathbf{x}}_{+,1}$
in the second period of the effective potential $\pi/k\leq x\leq 2\pi/k$.
Hence, for small values of $\gamma<\gamma_{\rm cr}$, the homoclinic bifurcation creates a saddle connection
(see, e.g., Refs.~\cite{H91,W90})
between $\bar{\mathbf{x}}_{+}$ and  $\bar{\mathbf{x}}_{+,1}$. In this connection, the branch of $W^s(\bar{\mathbf{x}}_{+})$ lying in the half-plane $y<0$ is -- simultaneously -- the branch of the unstable manifold $W^u(\bar{\mathbf{x}}_{+,1})$ lying also in the half-plane $y<0$.

When $\gamma<\gamma_{\rm cr}$ is further increased,
the branch of $W^s(\bar{\mathbf{x}}_{+})$ lying in the half-plane $y<0$,
connects to the unstable spiral point $\bar{\mathbf{x}}_{-,1}$ in the
second
period of the effective potential. Thus, another type of bifurcation occurs,
when $\gamma<\gamma_{\rm cr}$ exceeds a
threshold value $\gamma_{\mathrm{thr}}<\gamma_{\rm cr}$, namely, a {\it saddle connection breaking bifurcation}.
The existence of $\gamma_{\mathrm{thr}}$ is guaranteed by the fact that
as $\gamma< \gamma_{\rm cr}$ is increased, the growth and oscillation frequency
of the spiraling orbit far from the equilibrium $\bar{\mathbf{x}}_{+}$ dominates, enforcing the considered branch of
$W^s(\bar{\mathbf{x}}_{+})$ to form the shape of the spiral. Furthermore, in the regime $\gamma_{\mathrm{thr}}\leq\gamma<\gamma_{\rm cr}$, both branches of $W^u(\bar{\mathbf{x}}_{+})$
become unbounded.

When the bifurcation for $\gamma\geq\gamma_{\mathrm{cr}}$ occurs, the situation is exactly as in
the saddle connection breaking setting, with the difference that the unstable spirals are replaced by the
unstable nodes, and the orbits are escaping (converge) from the strip $Y_1\leq y\leq Y_2$
(on the stable manifolds) at an exponential rate.


We conclude with the case $\frac{E}{2A_k kV_0}=1$, where solely one fixed point $\bar{\mathbf{x}}$ exists, and the case where no-fixed points exist. For the former, which is a case of {\it saddle-node bifurcation}, the nullcline analysis
in a small neighborhood of $\bar{\mathbf{x}}$ reveals that it is unstable. For the latter,
the nullcline analysis in the region $B_0$, establishes
that any orbit must become unbounded.


\subsubsection{Phase-plane portraits}

\begin{figure}[tbp]
  \begin{center}
  \includegraphics[width=.23\textwidth]{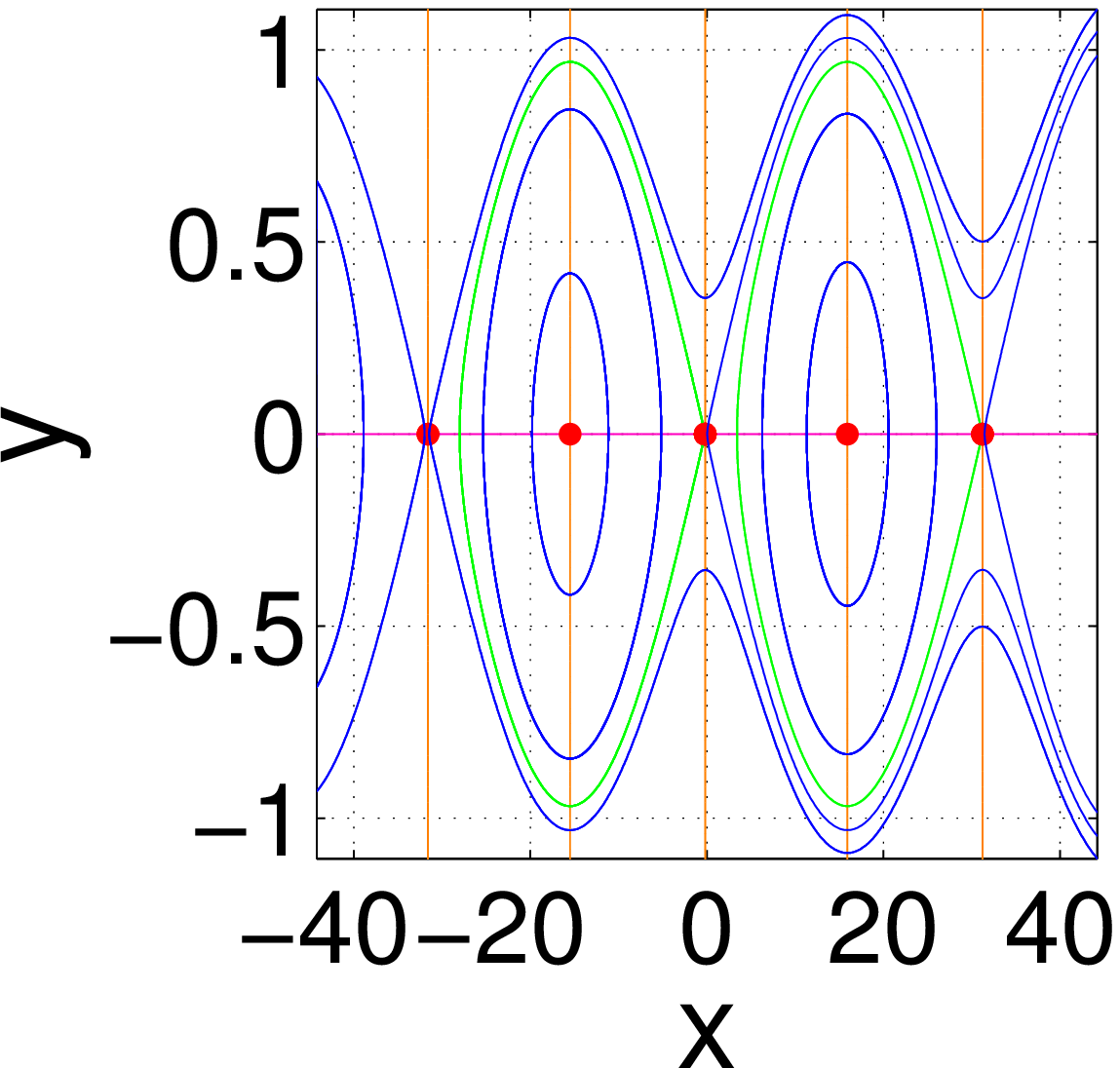}
  \includegraphics[width=.23\textwidth]{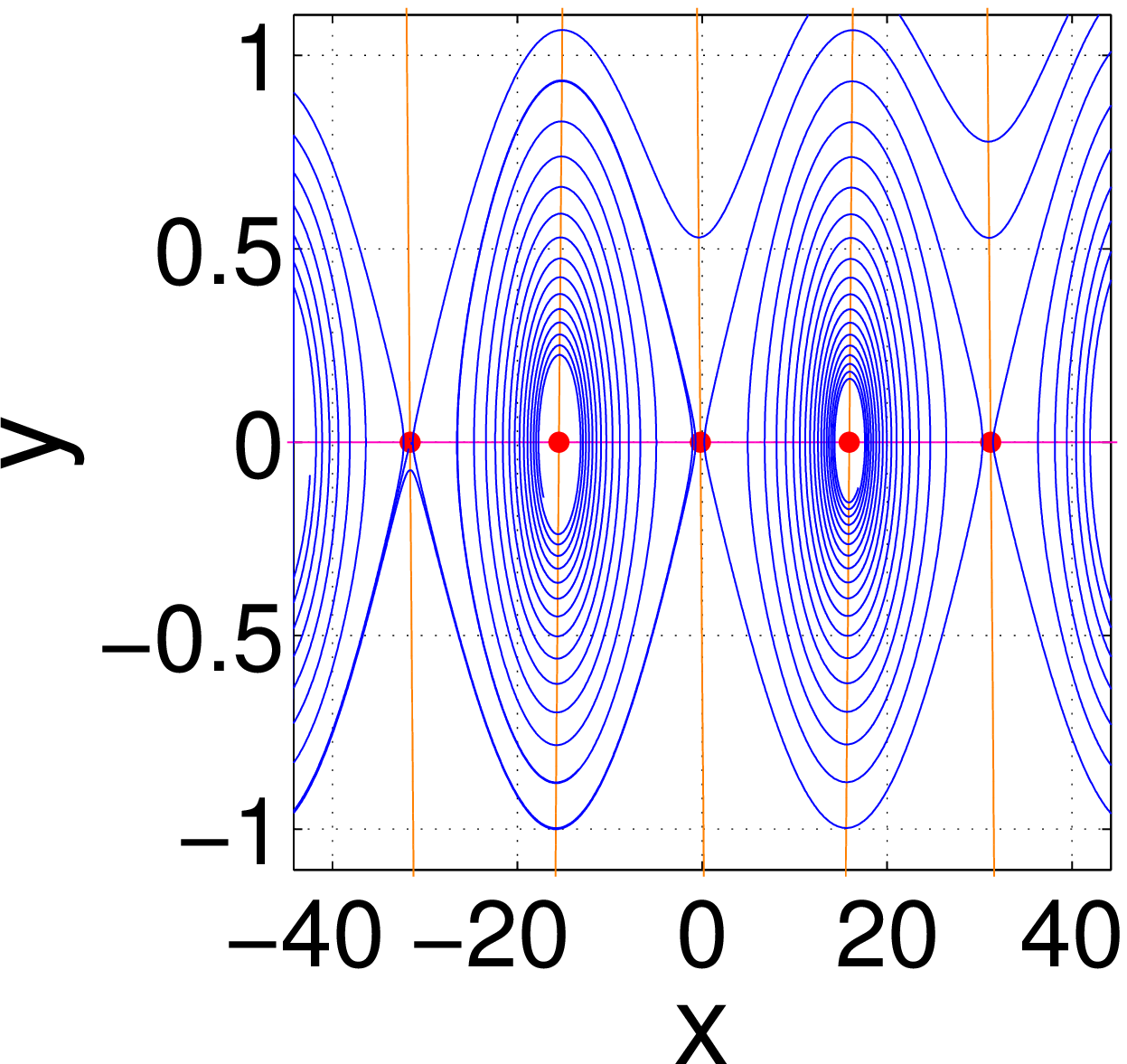}
  \includegraphics[width=.23\textwidth]{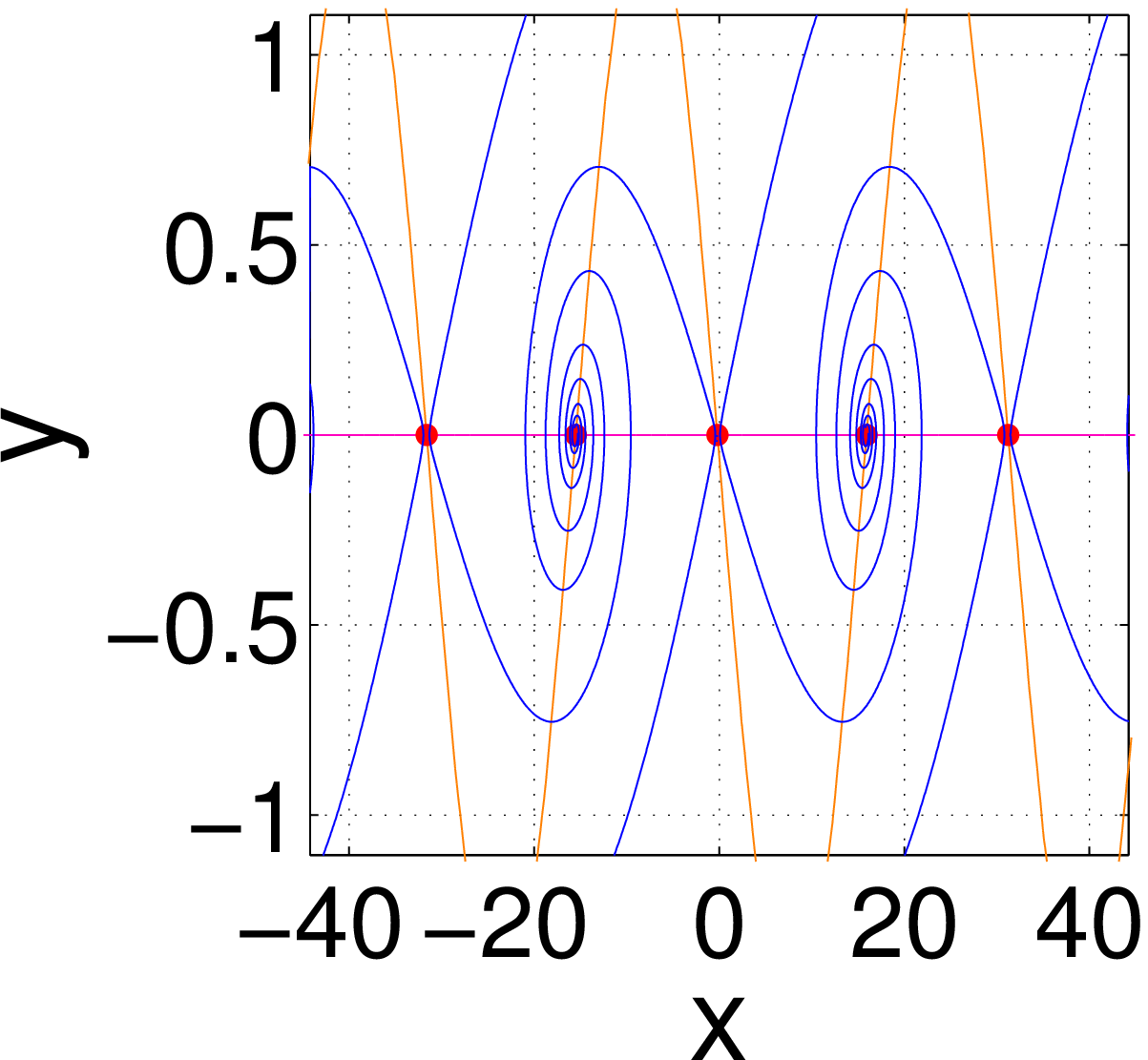}
  \includegraphics[width=.23\textwidth]{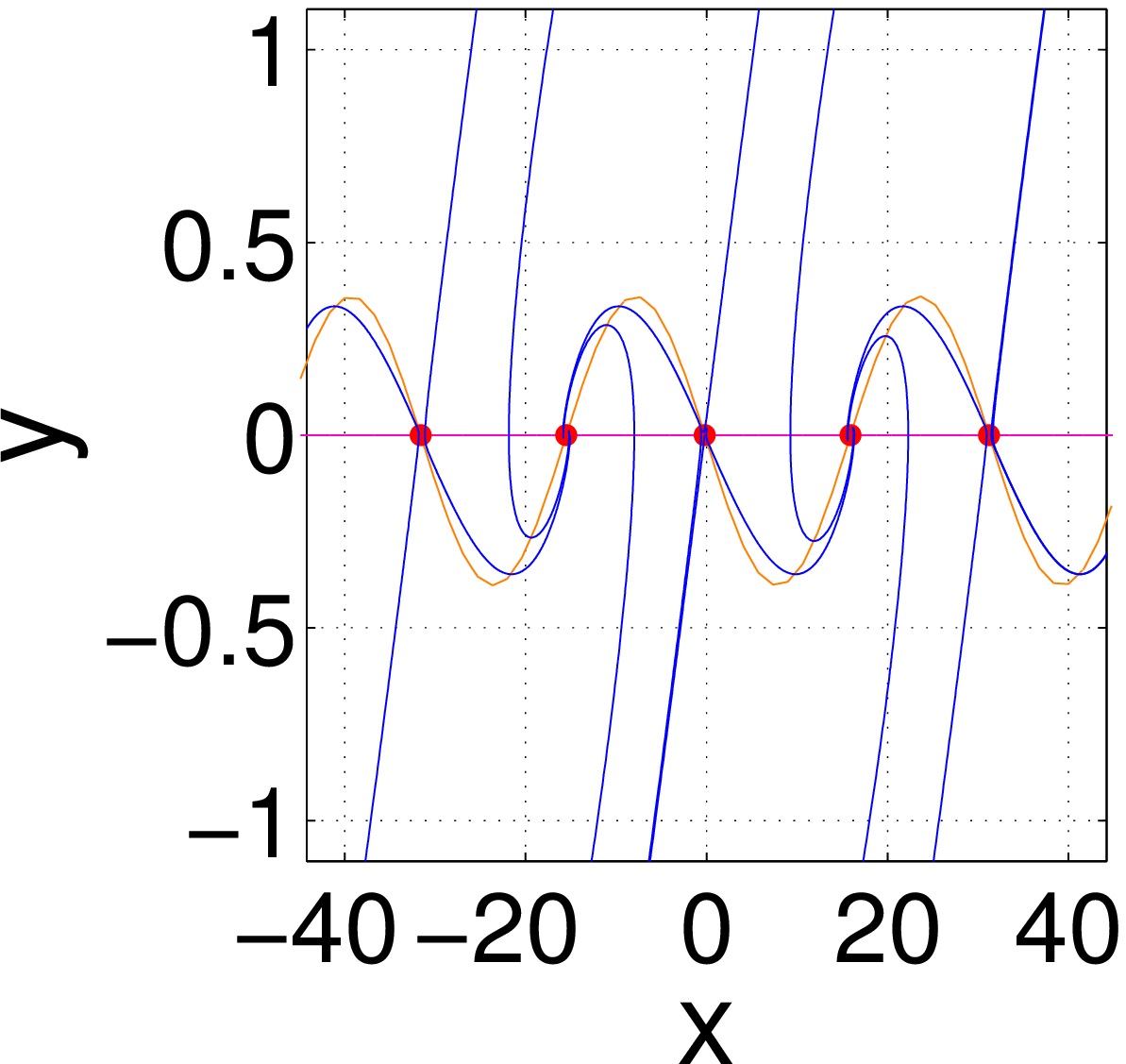}
  \end{center}
  \caption{(Color online) Phase planes for $\ E = 0.004, \ V_0 = 0.05,\ k=0.1,\ \mu=1$; the computational
  window is shown for two periods of the effective potential $U(x)$.
  Top left panel corresponds to the Hamiltonian case, $\gamma=0$, while for the top right, bottom left and bottom right panels, $\gamma=0.005$, $\gamma=0.05$ and $\gamma=0.2$, respectively. Bold (red) points depict the fixed points, and solid gray (orange) lines show the nullclines; the gray (green) lines in the bottom left panel depict the homoclinic orbits. }
  \label{pp}
\end{figure}

The above analysis on the structure of the phase space of the system (\ref{eqn_center_dark_system})
is illustrated in Fig.~\ref{pp},
for four different values of the damping parameter $\gamma$ (other parameters values are $k=0.1$, $E=0.004$, $V_0=0.05$ and $\mu=1$). The computational window is shown for two periods of the effective potential $U(x)$.

In particular, the top left panel presents the family of homoclinic orbits in the Hamiltonian case ($\gamma=0$), surrounding the stable centers. The non-homoclinic branch of $W^u(\bar{\mathbf{x}}_{+})$ diverges and becomes unbounded, while $W^s(\bar{\mathbf{x}}_{+})$ remains bounded.

In the top right panel, where the orbits have been plotted for $\gamma=0.005$, we observe the transition from purely imaginary eigenvalues to complex eigenvalues. The stable centers of the previous case, now become unstable spirals. Furthermore, we see the homoclinic bifurcation for the saddle. It is also interesting to observe the numerically computed orbits in the half-plane $y<0$ which
connect to the saddles, clearly illustrating that the system is in the regime of the saddle connection.

The bottom left panel corresponds to an increased value of $\gamma$, namely $\gamma=0.05$. In this case, the saddle connection breaking bifurcation has
already taken place. Now, the stable manifolds of the saddles consist of orbits which are coming from unstable spirals. The unstable manifolds, as well as all other orbits, become unbounded with increased growth, if compared with the previous case.

Finally, as shown in the bottom right-panel, the solutions grow exponentially fast in the case $\gamma=0.2$. The character of the instability has changed via
the collision of the two complex eigenvalues which have now become
real, and the unstable spirals have thus become unstable
nodes. The numerical computed phase plane demonstrates the particular
node-orbits joining the saddle points.

We note that above we have focused on the anti-damped case $\gamma>0$, which is physically relevant in the context of BECs. Nevertheless, the dynamical system (\ref{eqn_center_dark_system}) exhibits a rich behavior for $\gamma<0$, a case corresponding to a damped Josephson junction (see, e.g., Refs.~\cite{JJ1,JJ2,JJ3}). As an additional comment along this vein, suitable parallels between the
two cases can be drawn on the basis of the invariance of
Eq.~(\ref{eqn_center_original}) under the transformation
$\gamma \rightarrow -\gamma$ and $t \rightarrow -t$.

%

\subsection{Bogoliubov-de Gennes (BdG) analysis}

We complement this dynamical systems picture by briefly describing our methods for obtaining
the numerical results for the DGPE (see next section) and comparing them with our analytical predictions.
First, we will identify the fixed points of the dynamical system of Eqs.~(\ref{eqn_center_dark_system})
-- corresponding to stationary dark soliton states
of the DGPE, Eq.~(\ref{GP}) -- through a Newton-Raphson-type iteration method.
Subsequently, we will study the linear stability of the stationary solutions
of the DGPE
by means of the following ansatz:
\begin{equation}
u(x,t)=u_0(x)+\epsilon [a(x)e^{\lambda t}+b^*(x)e^{\lambda^* t}].
\label{bdg}
\end{equation}
Here, $u_0(x)$ is the stationary solution of
Eq.~(\ref{GP}) and the term proportional to
$\epsilon$ (which is a formal small parameter)
represents a small perturbation.
Substituting
the above expression into Eq.~(\ref{GP}), and linearizing with respect to the
amplitudes $a$ and $b^*$, we obtain at $\mathcal{O}(\epsilon)$
the following linear [Bogoliubov-de Gennes (BdG)] stability equations:
%
\[ \left( \begin{array}{ccc}
L_1 & -u_0^2 \\
{u_0^*}^2 &L_2  \end{array} \right)
 \left( \begin{array}{ccc}
a \\
b \end{array} \right)=\lambda \left( \begin{array}{ccc}
-i+\gamma & 0 \\
0 &-i-\gamma  \end{array} \right) \left( \begin{array}{ccc}
a \\
b \end{array} \right)\]
%
where $L_1 = \frac{1}{2}
\partial_x^2 -2|u_0|^2+\mu-V$ and
$L_2 = -\frac{1}{2}
\partial_x^2+2|u_0|^2-\mu+V $.
%
This way, the linear stability problem is transformed
into a spectral problem for the linearization operator:
\[L = \left( \begin{array}{ccc}
\frac{L_1}{-i+\gamma} & \frac{-u_0^2}{-i+\gamma} \\
\frac{{u_0^*}^2}{-i-\gamma} &\frac{L_2}{-i-\gamma}  \end{array} \right).
\]
The latter operator is discretized (on the same grid in which the solution
is computed) and its eigenvalues $\lambda$ and eigenvectors
$\{a(x),b(x)\}$ are computed numerically.

\begin{figure}[tbp]
  \begin{center}
    \includegraphics[width=0.23\textwidth]{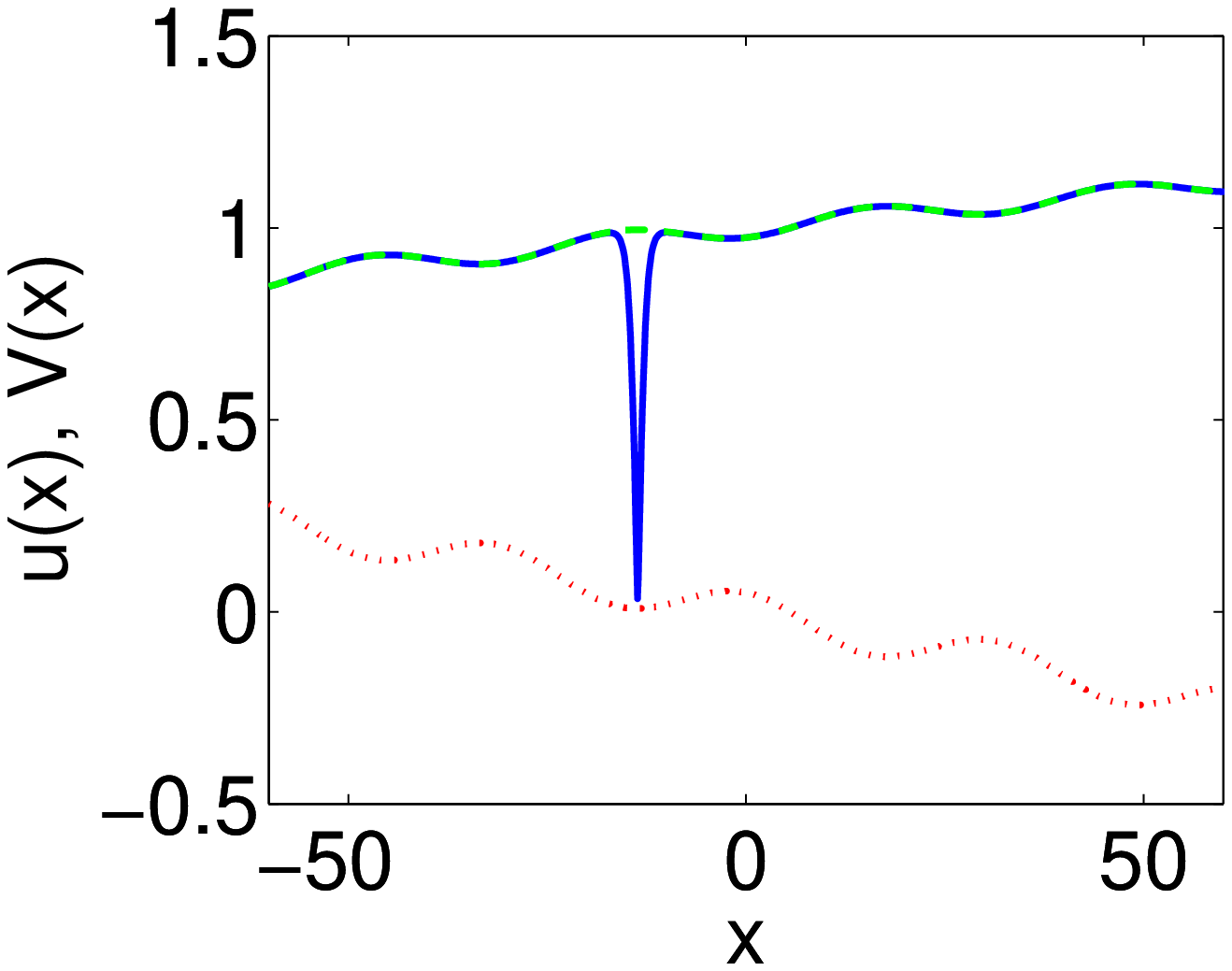}
    \includegraphics[width=0.23\textwidth]{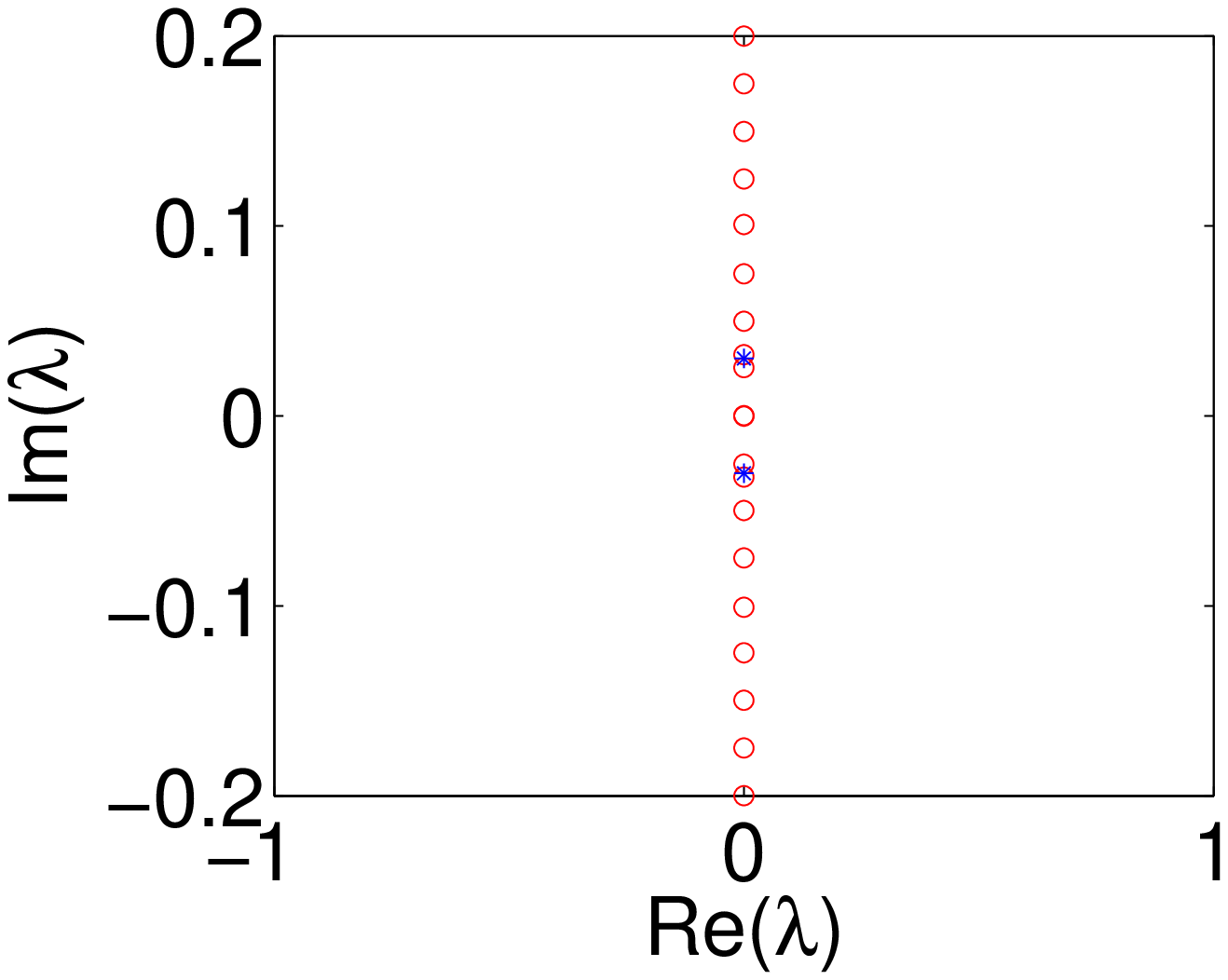}
    \includegraphics[width=0.23\textwidth]{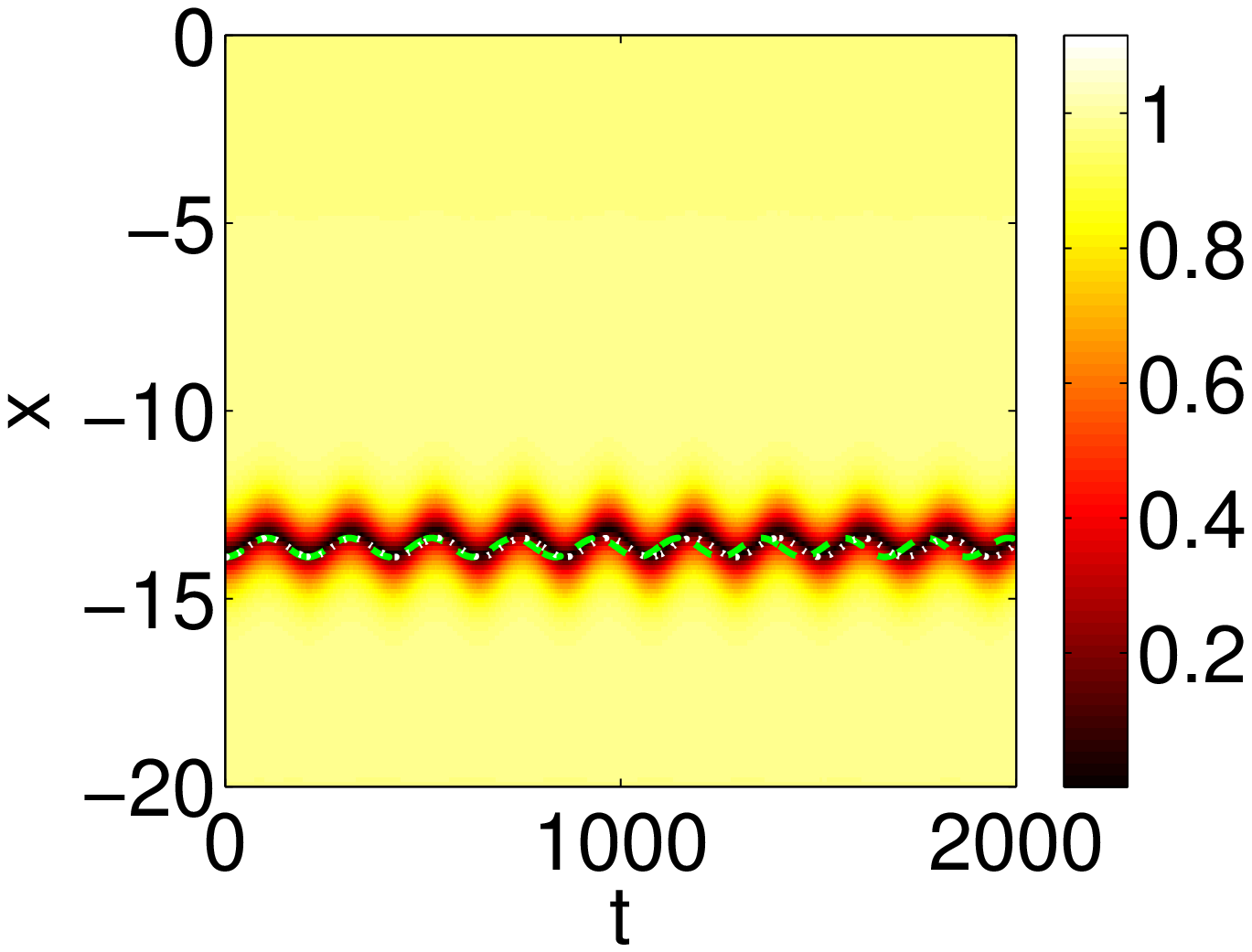}
    \includegraphics[width=0.23\textwidth]{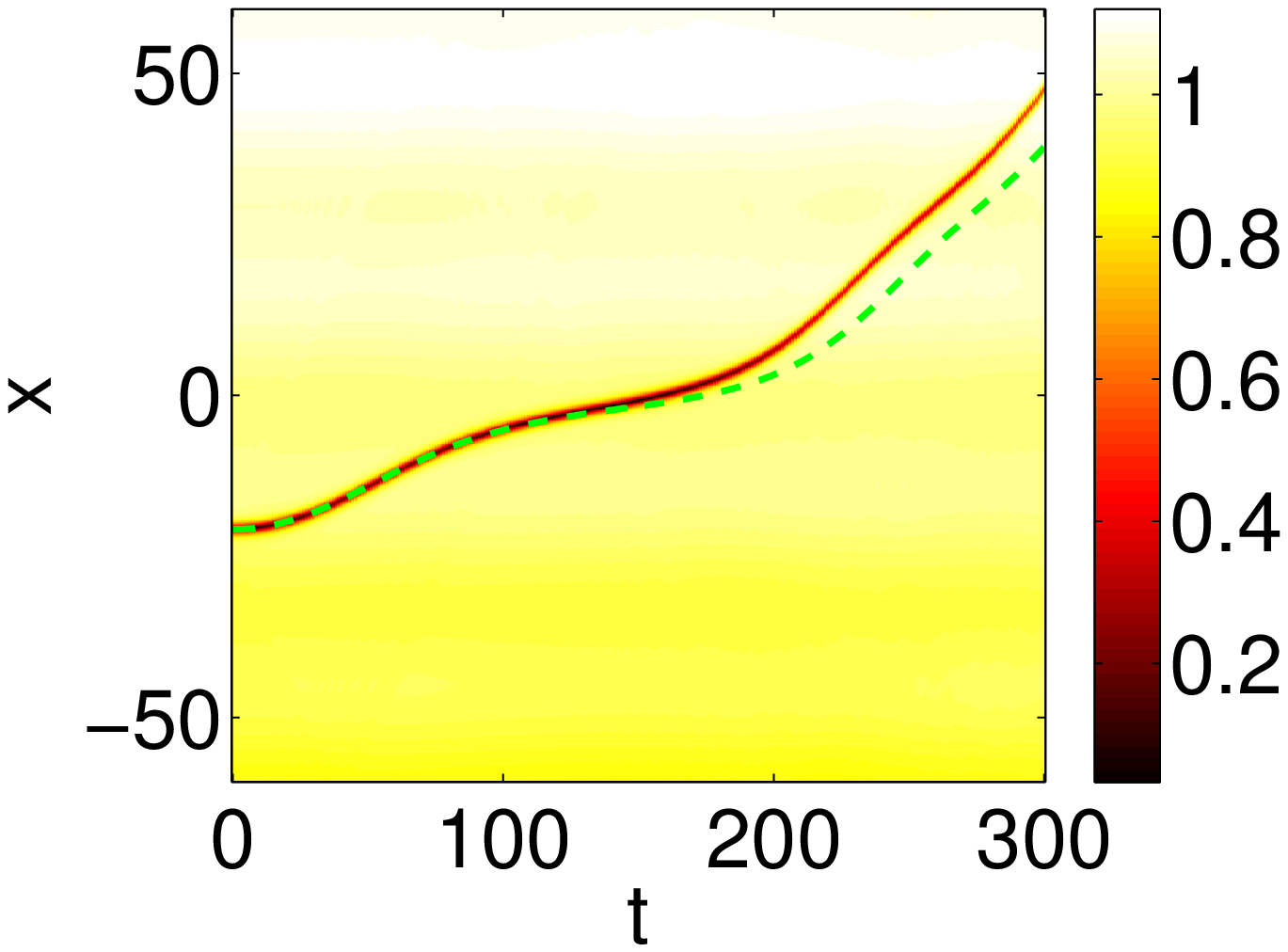}
    \end{center}\caption{
(Color online)
The top left
panel shows the stationary solution in (blue) solid line with the dark soliton
at the local minimum of the external potential shown in (red) dotted line; the (green) dashed
line shows the background. The
top right
panel shows the corresponding eigenvalue of
the linear stability problem, with the (blue) star indicating
the theoretical eigenfrequency (and the (red) circles illustrating the 	
DGPE result). The
bottom left
panel shows the dynamics of the DGPE
with the dark soliton initially located at $x_0=-13.89$; the (green) dashed
line is the evolution of the ODE.
The bottom
right
panel shows the DGPE dynamics
with a dark soliton initially
placed at $x_0=-20.81$ (i.e., near the turning
point for the homoclinic orbit), while the (green) dashed line shows
the corresponding dynamics of the ODE.
Parameter values are: $\gamma=0,\ E = 0.004, \ V_0 = 0.05,\ k=0.1,\ \mu=1$. }
  \label{gamma_0_E_0004_V0_005_k_01_mu_1_min}
\end{figure}

In the following section we will first identify
the fixed points
in the partial differential equation (PDE) of Eq.~(\ref{GP}),
and in the (anti-damped Josephson junction) equation of motion of the dark soliton center, namely the
ordinary differential equation (ODE) of  Eq.~(\ref{eqn_center_original}).
The result of the fixed point scheme convergence for the PDE
equilibrium will be
compared to the solution of the transcendental
equation for the dark soliton center [cf. Eq.~(\ref{tran})].
Then, we will
compare the eigenvalue problem of the
linearized ODE [i.e., the explicit expression for the eigenvalues of
the Jacobian of Eq.~(\ref{jac})]
and that of the full DGPE (i.e., the pertinent eigenvalues
of the operator $L$). Finally, we will proceed to
an even more demanding comparison, namely that of the full
dynamics of the infinite degrees of freedom system (the DGPE) and that
of its one degree of freedom reduction (the
ODE for the dark soliton center).
Both of these evolutions will be followed by means of a time-stepping
scheme (namely an explicit fourth-order Runge-Kutta method) and the
results will be illustrated, for direct comparison, in the same plot.
%

\section{Numerical results}


First, let us offer a few technical remarks, before we commence the presentation of our
results. We have used no flux boundary conditions (as periodic and Dirichlet
ones seem less appropriate for our setting). Also, in our numerical
continuations for the exact solutions and their linear stability, we have
always commenced from a well-known, benchmark limit (e.g., the one without
potential or without the thermal effects) and subsequently performed
parametric continuations in parameters such as typically $E$, $V_0$
and $\gamma$. When solutions were found to be unstable, the dynamical
evolution of the instability was monitored through the time dynamics.
Our principal results can be summarized as follows.

\subsection{The Hamiltonian case}

\subsubsection{Solitons located at the center fixed point}

\begin{figure}[!h]
  \begin{center}
    \includegraphics[width=0.23\textwidth]{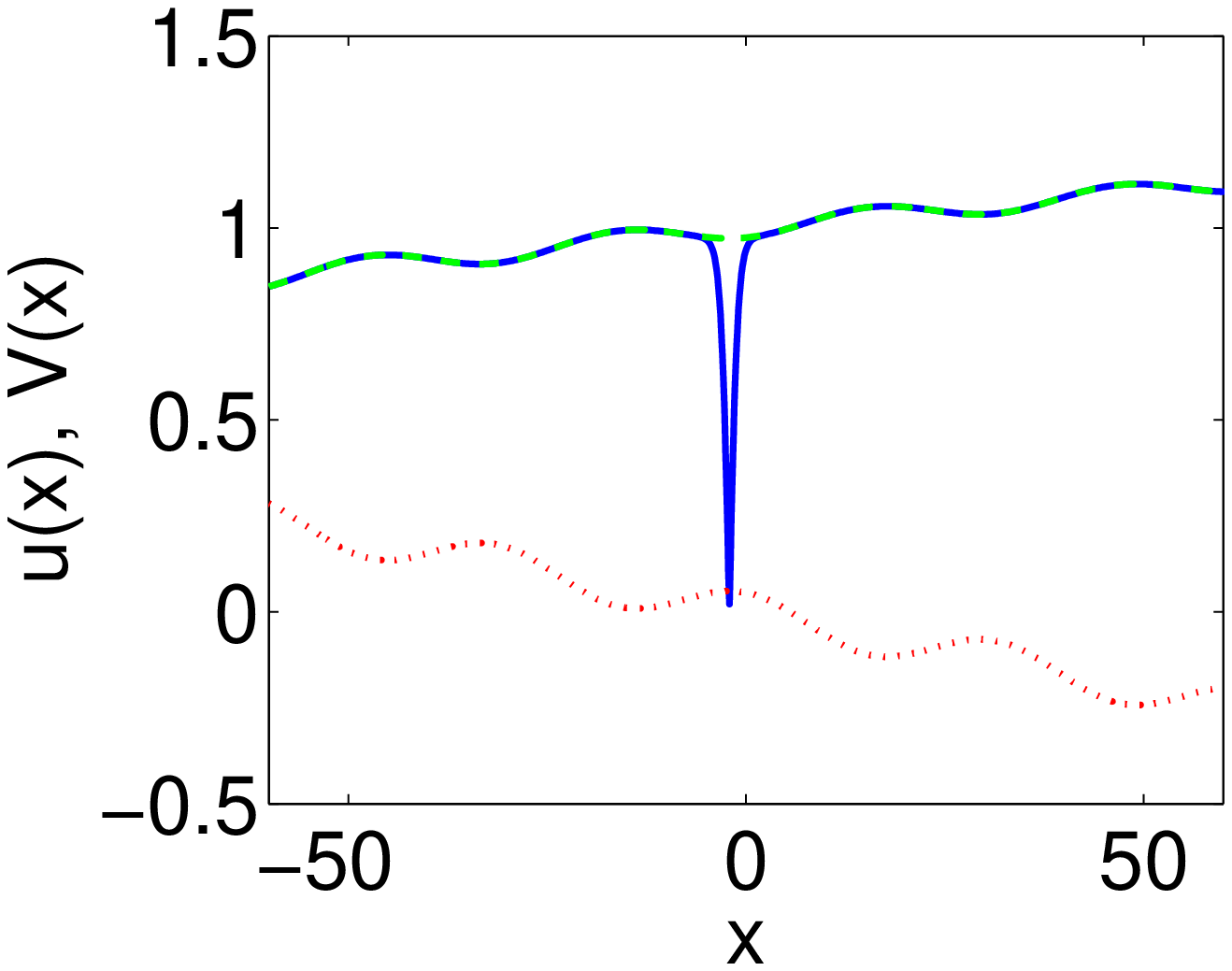}
    \includegraphics[width=0.23\textwidth]{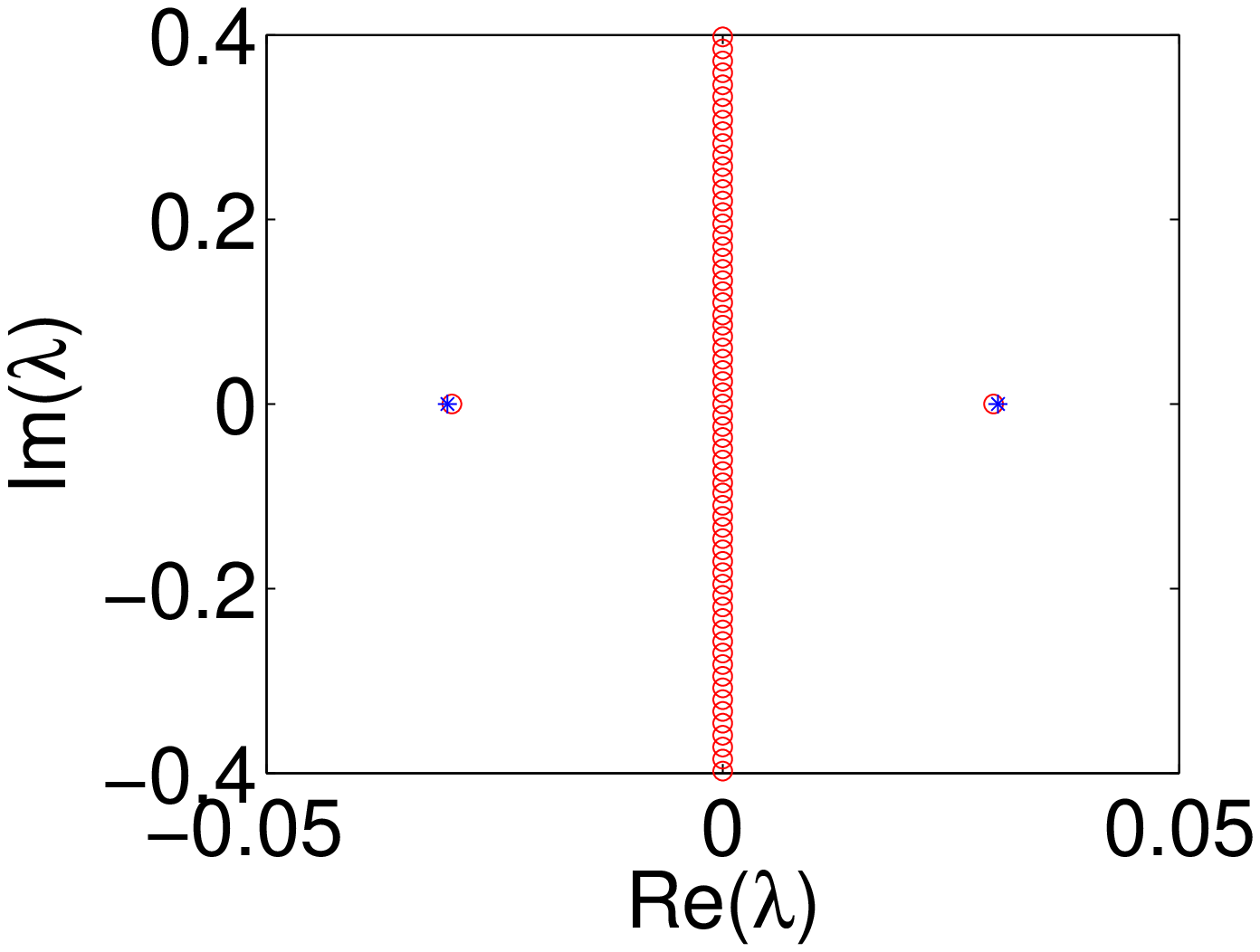}
    \includegraphics[width=0.23\textwidth]{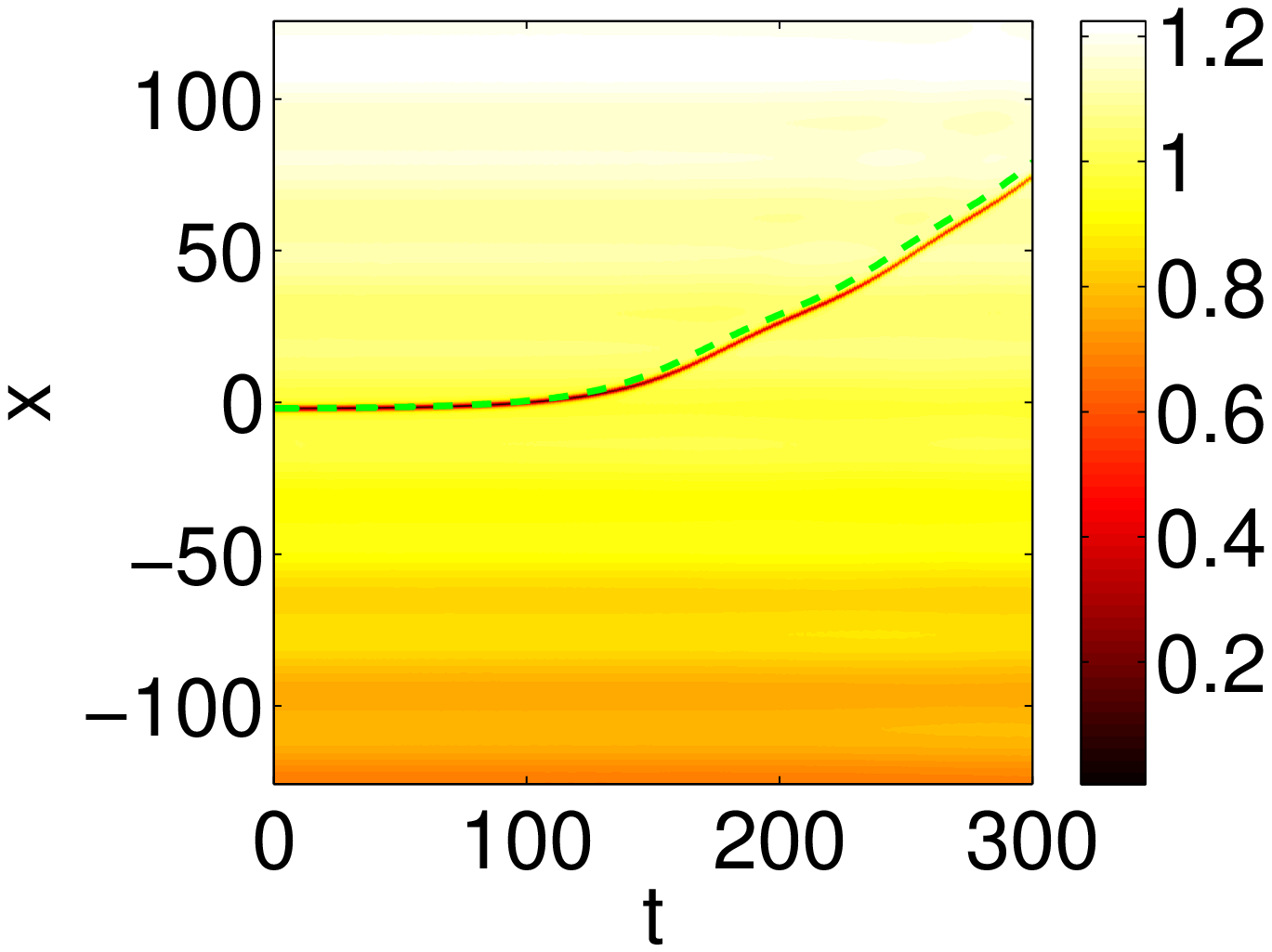}
    \includegraphics[width=0.23\textwidth]{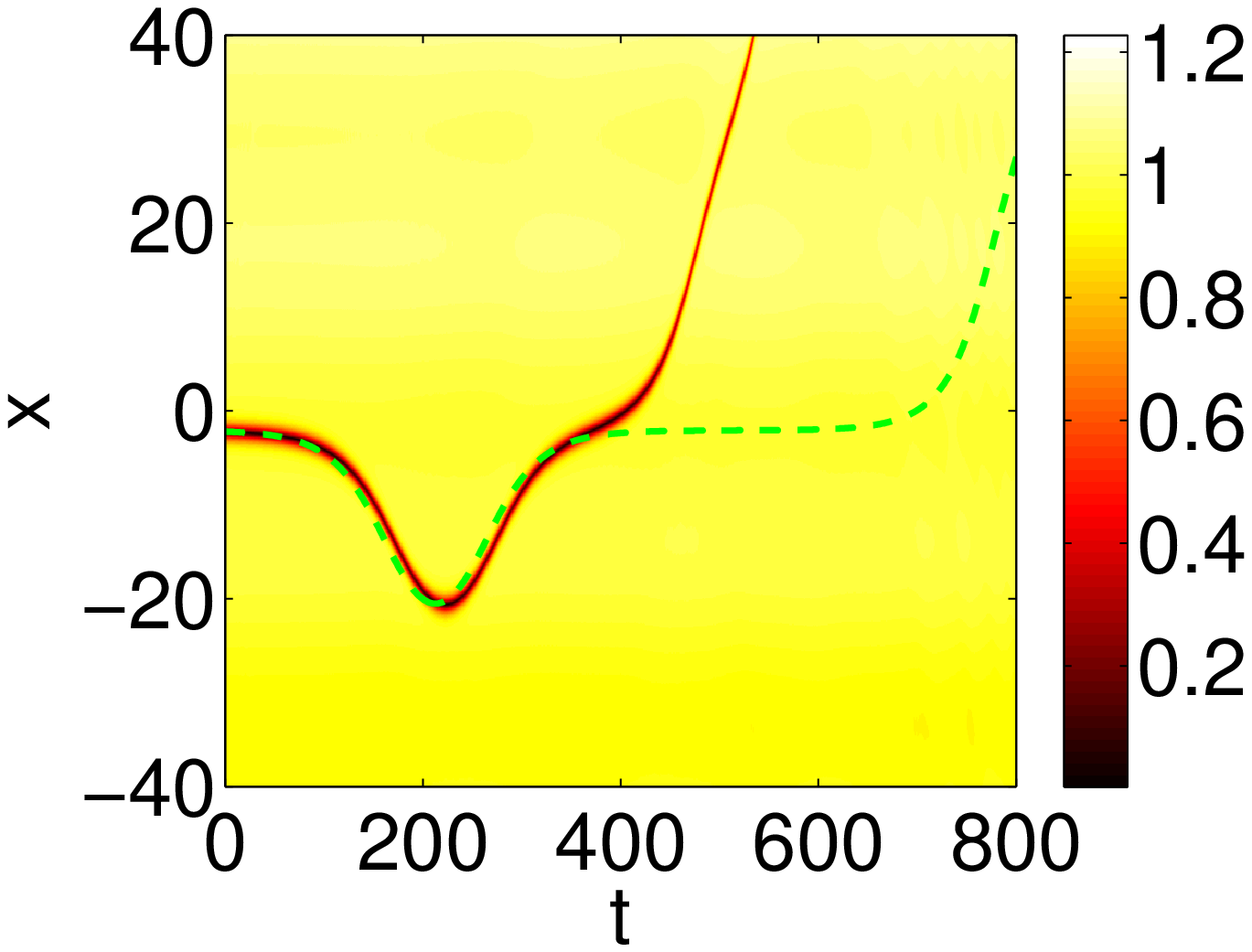}  \end{center}\caption{(Color online) The top left
    panel shows the stationary solution with
the dark soliton at the maximum of the effective potential.
The top right
panel shows the corresponding eigenvalues
of the linear stability analysis;
Again, in addition to the (red) circles stemming from the eigenvalues
of the operator $L$, there exists
the theoretically predicted growth rate [indicated by the (blue) star]
from the corresponding ODE. The two bottom panels show the dynamics of the
DGPE, with
the dark soliton initially placed at the local maximum of
the effective potential, with two different perturbations:
the difference refers to the sign of the
unstable eigenvector of the linearization problem, which has been
added to the unstable dark soliton in order to initiate
the instability evolution. The amplitude used in the two cases
shown is $\pm0.1$. Parameter values are: $\gamma=0,\ E = 0.004, \ V_0 = 0.05,\ k=0.1,\ \mu=1$. }
  \label{gamma_0_E_0004_V0_005_k_01_mu_1_max}
\end{figure}

In the Hamiltonian case, i.e., for $\gamma=0$,
the phase plane of Eq.~(\ref{eqn_center_dark_system}) is shown in the
top panel of Fig.~\ref{pp}
(other parameter values are: $E = 0.004, \ V_0 = 0.05,\ k=0.1,\ \mu=1$).
Direct inspection of the phase plane suggests that the center of the dark soliton
may periodically oscillate around the (center)
fixed point $(\bar{x}_-,0) = (-13.6363,0)$ with a linearization frequency
(of small amplitude oscillations) of $\omega_{\rm eff}=\sqrt{\alpha}=0.0302$
(recall that $\alpha= k\sqrt{4A_k^2k^2V_0^2-E^2}$).
On the other hand, there is a homoclinic orbit passing through the saddle
$(\bar{x}_+,0)=(-2.0709,0)$ which  intersects the $x$-axis at (a finite turning
point of) about $-20.5$.
This phenomenology at the level of the DGPE (and the comparison with the
ODE) is shown in Fig.~\ref{gamma_0_E_0004_V0_005_k_01_mu_1_min}.
In the top left
panel of the figure, the solid (blue) line shows an example
of the exact stationary solution at the effective potential minimum as
obtained from our fixed-point iteration method. Furthermore,
the dotted (red) line denotes the external potential, while
the dashed (green) line is the the background $u_b(x)$, i.e., the ground
state of the system. The latter showcases the validity of our ansatz
approximation assuming that the dark soliton can be embedded in the
system through a product ansatz, where it is directly multiplied to the
(back)ground state of the system.
The top right
panel of the figure presents a comparison of the
eigenvalues of the matrix $L$ which are found to be
purely imaginary, corroborating the neutral stability of this
center point. Within this plot, the (blue) star represents
the theoretical frequency $\pm 0.0302$, which nearly
coincides with one of the eigenvalues $\pm 0.0292i$ of $L$.
As has been analyzed in
detail in previous works (see, e.g., Refs.~\cite{review,andreas,gt07}) for
the Hamiltonian case, and even for the thermal BEC
setting~\cite{Cockburn,cock2}, this eigenvalue represents the
so-called anomalous (or negative energy, or negative Krein
signature)~\cite{review} mode, which is the one that
pertains to the (eigenfrequency of the) motion of the dark
soliton around its equilibrium
position. As seen in the
bottom left
panel of Fig.~\ref{gamma_0_E_0004_V0_005_k_01_mu_1_min},
the (white) dotted line oscillating with the theoretical
frequency $0.0292$ provides a quite accurate description
of the full dynamics of the DGPE around its center equilibrium position
(the
soliton is
initially placed at $x_0=-13.89$),
confirming our expectation that it is indeed this eigenmode that
describes the
dark soliton motion.
Nevertheless, we have subjected the ODE approximation to a far more
stringent test by initializing the
soliton at $x_0=-20.81$, namely
very near the turning point for the homoclinic orbit (i.e., the most
unstable orbit of the system). However, even in that case as attested
by the bottom right panel of the figure, the agreement
between the ODE and PDE results
is quite striking.

\begin{figure}[tbp]
  \begin{center}
    \includegraphics[width=0.23\textwidth]{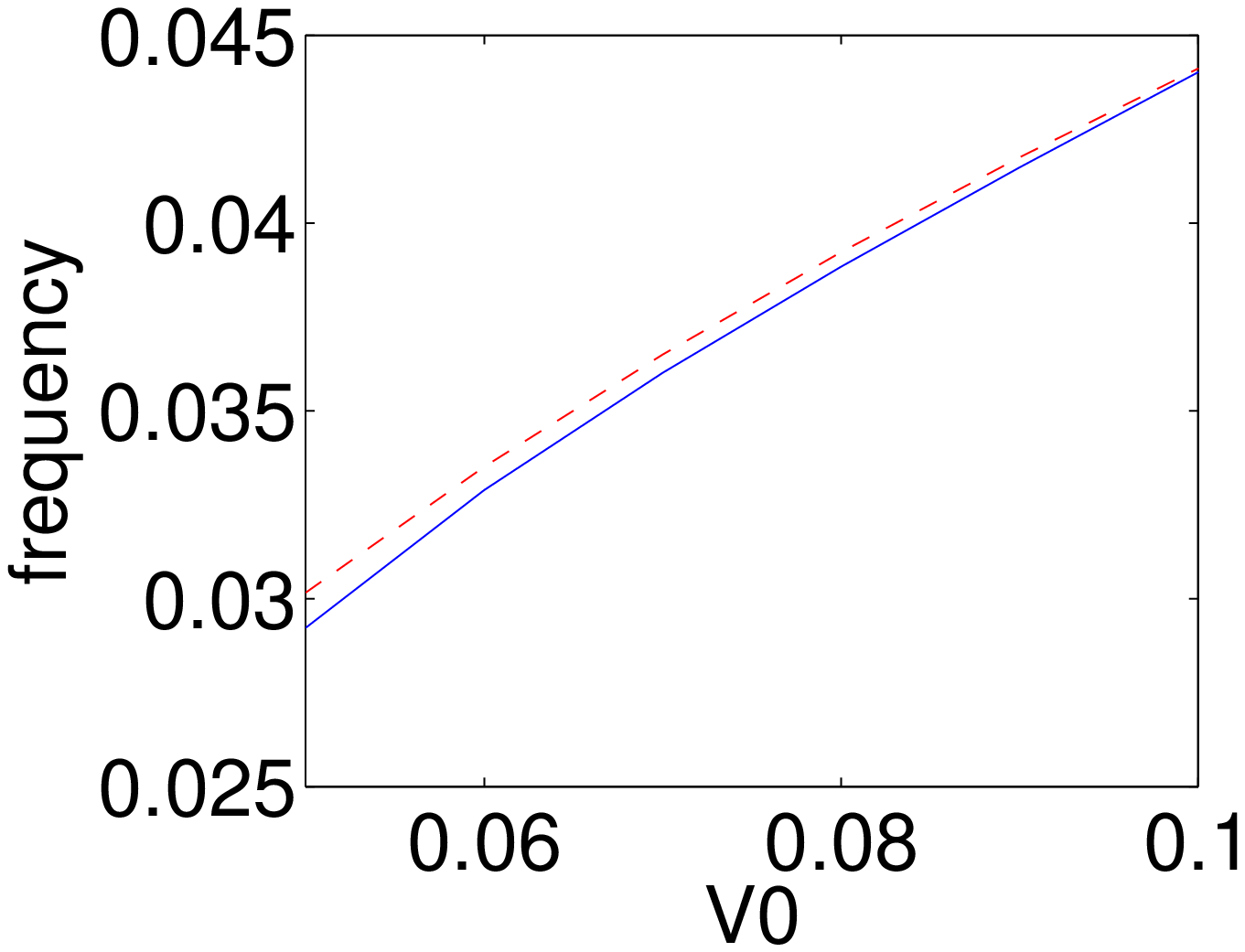}
    \includegraphics[width=0.23\textwidth]{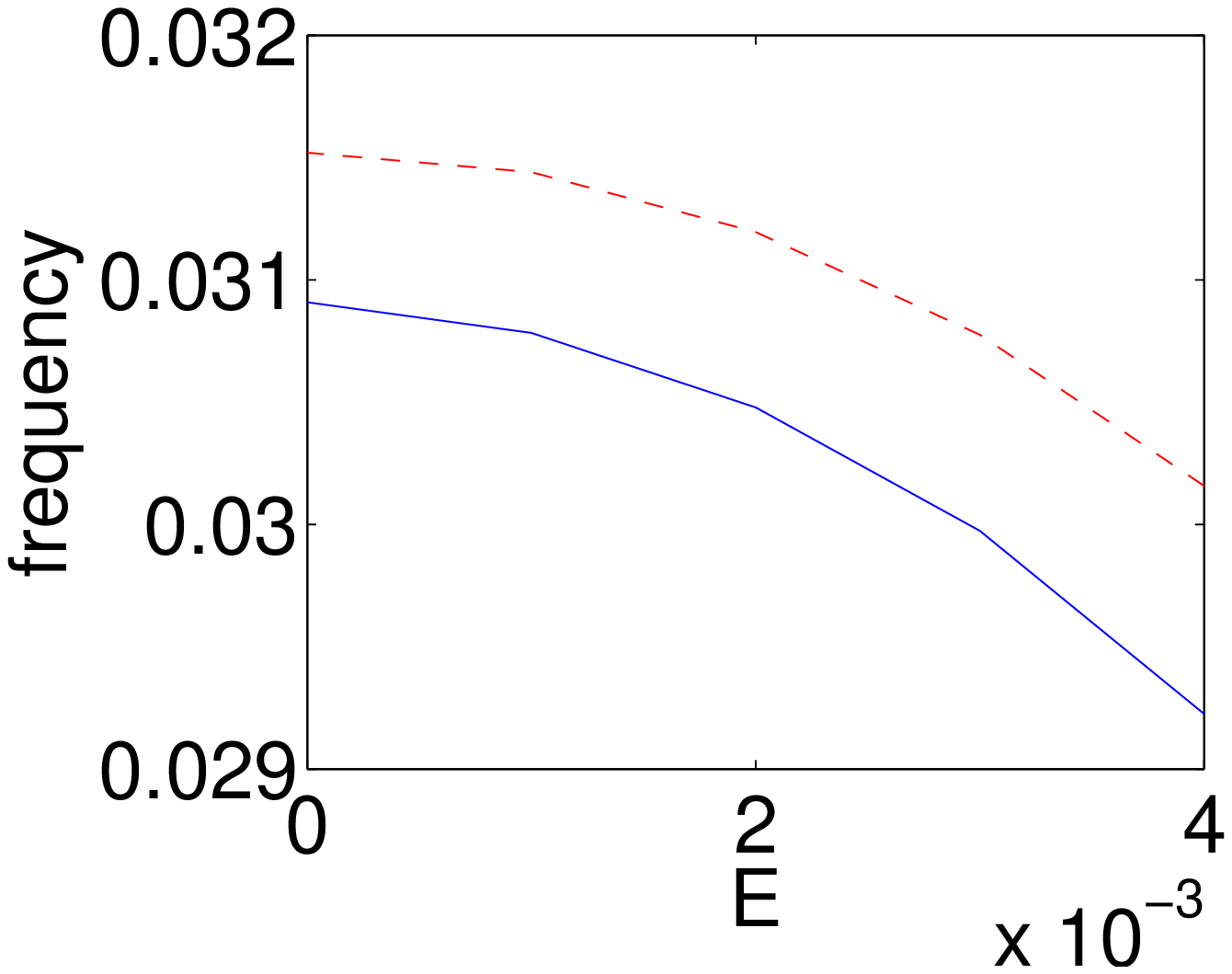}
    \includegraphics[width=0.23\textwidth]{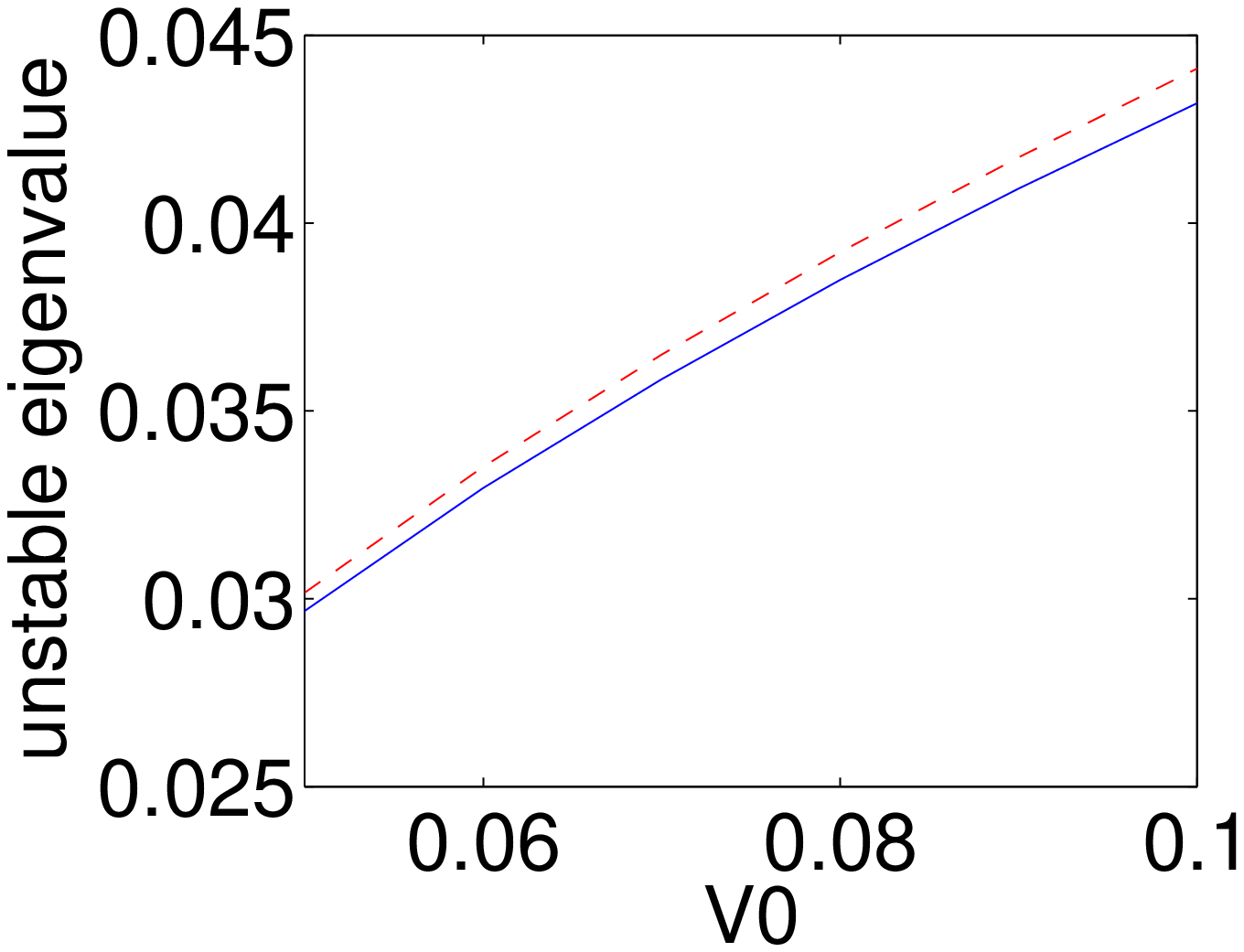}
    \includegraphics[width=0.23\textwidth]{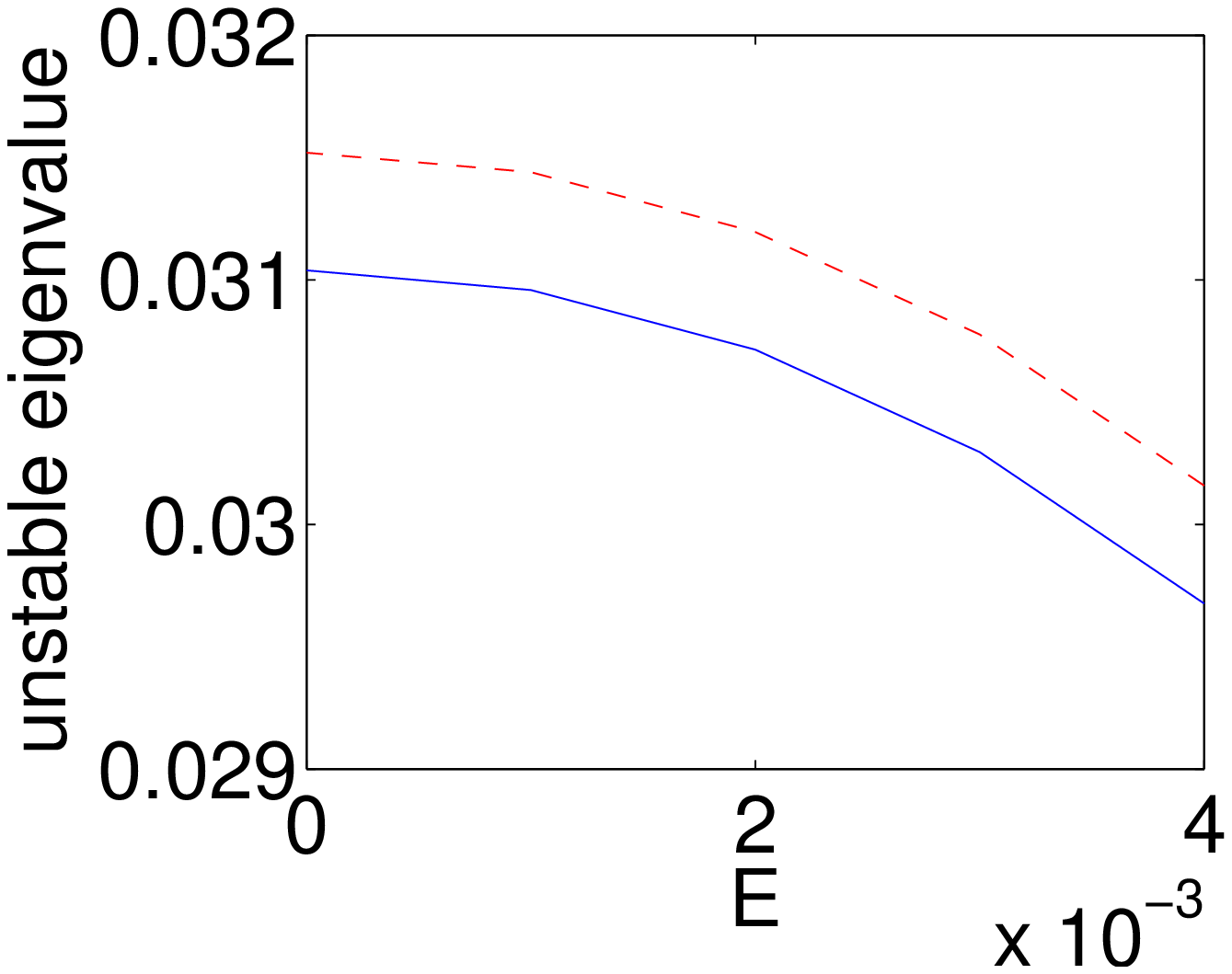}
  \end{center}\caption{(Color online)
The top and bottom sets of panels are for the
dark soliton at the local minimum and maximum of the effective potential, respectively. The
left
panels
showcase the change of frequency (or unstable eigenvalue)
vs.~$V_0$ for $E=0.004$, while the right ones illustrate
the change of frequency (or unstable eigenvalue)
vs.~$E$ for $V_0=0.05$. The (red) dashed
line is the result obtained analytically from the ODE,
while the solid (blue) line represents the corresponding mode
within the spectrum of linearization around the dark soliton of the DGPE.
Parameter values are: $\gamma=0,\ k=0.1,\ \mu=1$.}
  \label{gamma_0_frequency}
\end{figure}

\subsubsection{Solitons located at the saddle fixed point}

We now turn to an examination of the unstable saddle
fixed point existing at the local maximum of the effective potential, as illustrated in
Fig.~\ref{gamma_0_E_0004_V0_005_k_01_mu_1_max}.
The relevant solution is again shown in the top left panel, while,
this time, the linear stability problem of the DGPE possesses a real
pair of eigenvalues illustrating its saddle character.
These are again in reasonable agreement with the expectation of the
perturbation theory growth rates, as shown in the top right panel of the figure.
In order to show different possibilities of evolution
associated with this saddle point, we perturb its unstable eigenvector
with a positive amplitude and with a negative amplitude, respectively,
in the two bottom panels of Fig.~\ref{gamma_0_E_0004_V0_005_k_01_mu_1_max}.
In the case of the bottom left, the dark soliton acquires a velocity towards
positive values of $x$ and thus escapes to infinity sliding downhill
(i.e., washboarding) through the subsequent potential wells.
In the case of the right panel, the opposite
sign of the perturbation forces the soliton to move in the opposite direction,
where it encounters a turning point. However, after hitting the turning
point and changing direction of motion, it has enough energy to overcome
the local energy maximum (at the saddle point) and also
washboard towards $+\infty$. In both cases, we find that the ODE
dynamics captures fairly accurate the observations of the DGPE.
Nevertheless, in the right one the dynamics is very close to the
homoclinic orbit and hence is most unstable, a feature which is observed
with the longer meandering of the ODE trajectory in the vicinity of
the saddle point before it eventually escapes towards $+\infty$. It should
be noted that this feature
should perhaps naturally be expected:
in the infinite degrees of freedom system (the PDE), the
soliton has the ability to
shed (weak, but important in this case) radiative wavepackets, whereby
it becomes shallower and faster and, hence, it is easier for it to overcome
the relevant barrier. On the other hand, such decay channels are
forbidden by virtue of the ansatz in the one degree of freedom reduction.



Having captured an individual example of both types of fixed points,
we illustrate in Fig.~\ref{gamma_0_frequency} the results of
a corresponding continuation over the two parameters, the
strength $E$ of the linear (electric or gravitational)
potential and that
of the periodic (OL) potential, $V_0$, in both cases for $\gamma=0$ i.e.,
in the Hamiltonian case.
As predicted by the perturbation theory, when the dark soliton is around the local min (max) of the effective potential,
the frequency (growth rate) of the movement of dark soliton increases
as $V_0$ increases, while it decreases as $E$ increases. Both of these
dependencies are essentially encapsulated in the functional form of
$\alpha$ (since the relevant eigenvalue is predicted to be
$\pm i \alpha$ in the top panel, while should be $\pm \alpha$ in the bottom one.

%

\subsubsection{On the accuracy of perturbation theory}

\begin{figure}[tbp]
  \begin{center}
    \includegraphics[width=0.23\textwidth]{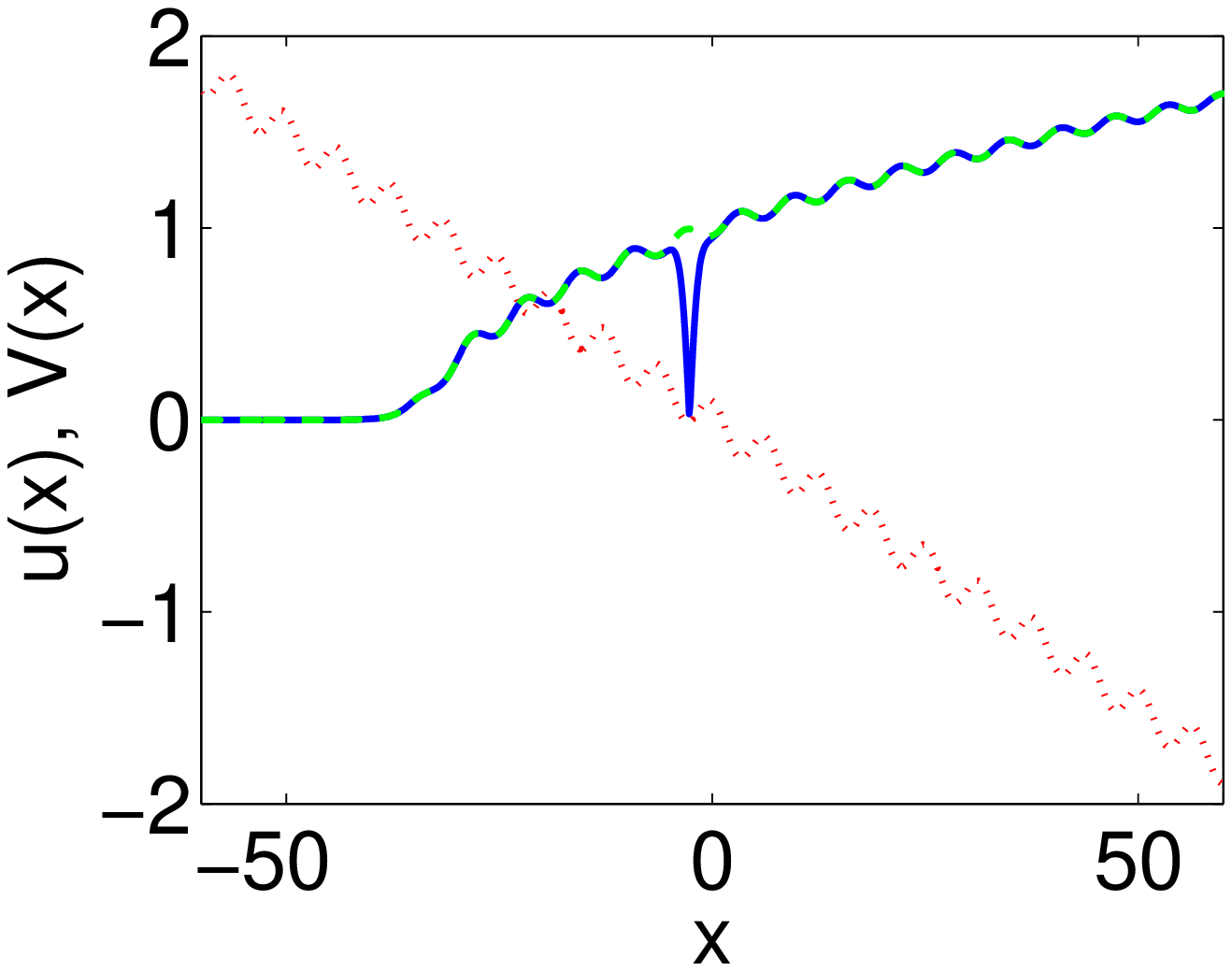}
    \includegraphics[width=0.23\textwidth]{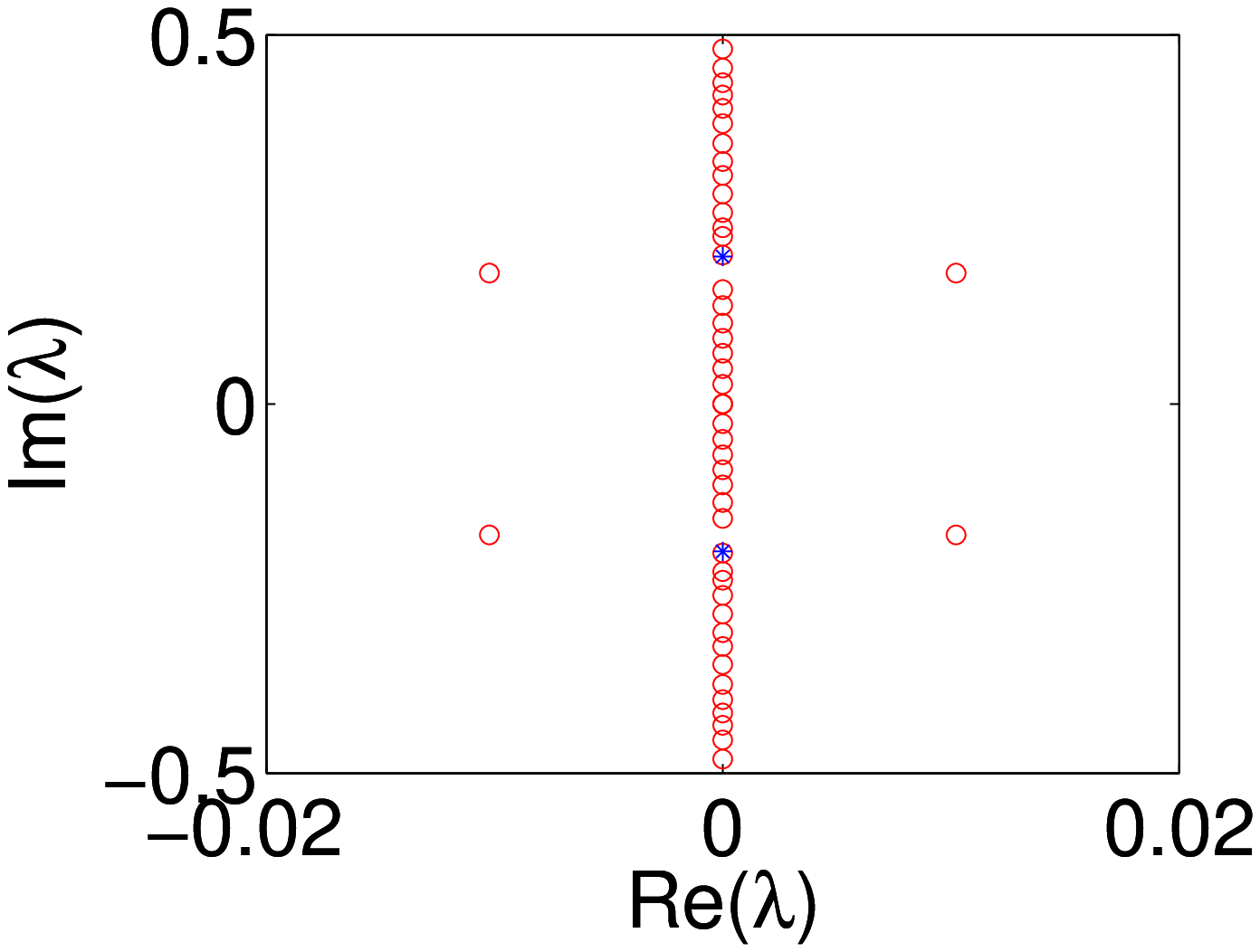} \\
    \includegraphics[width=0.33\textwidth]{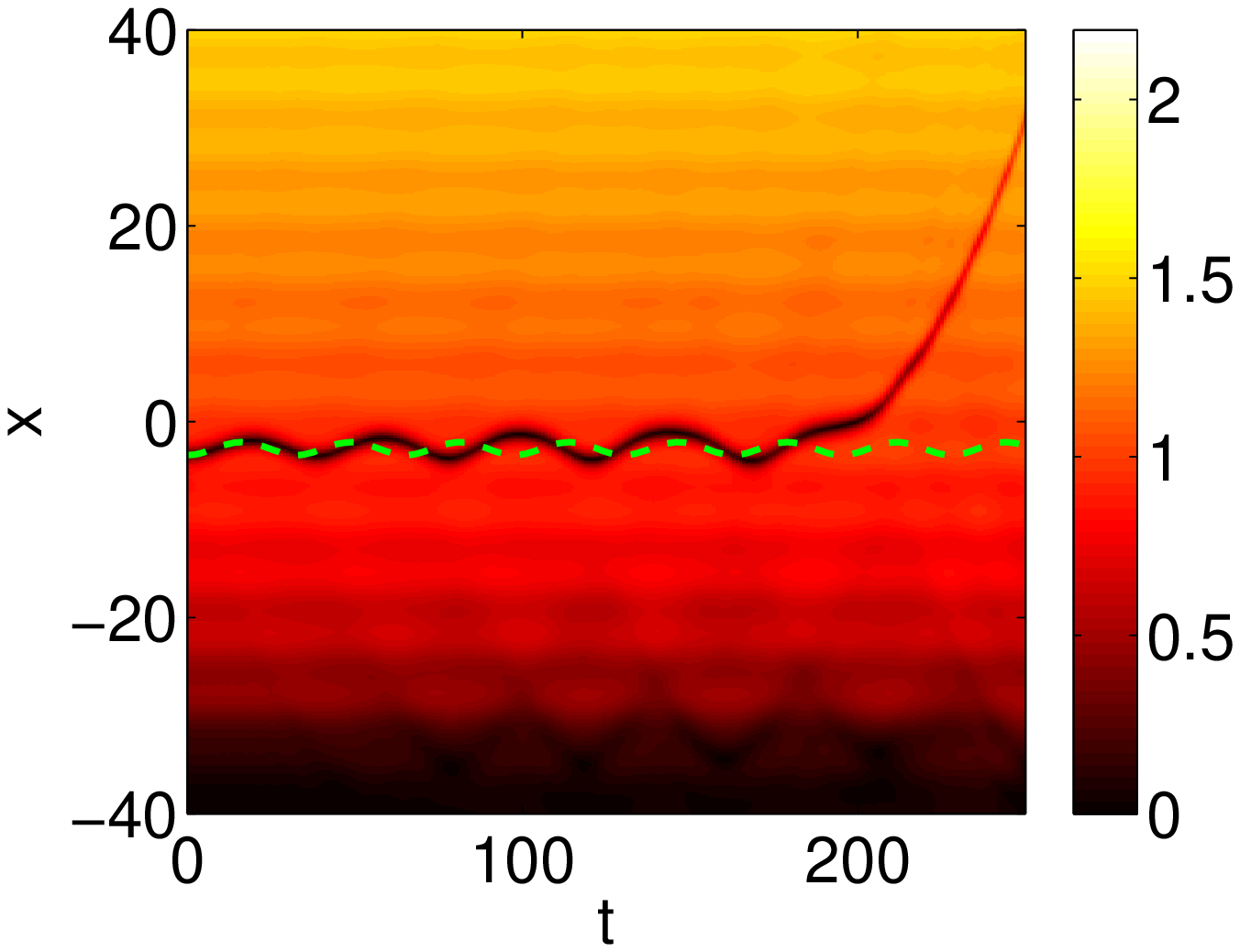}
  \end{center}\caption{(Color online) The top
  left
  panel shows the stationary dark
  soliton located at the local minimum of the effective potential. The
  top right panel shows its corresponding eigenvalues of the linear stability problem, which
indicate its oscillatory instability. The
bottom panel shows the
full dynamics of Eq.~(\ref{GP}) in its underlying contour plot
with the dark soliton initially at
$x_0=-3.41$ (around the center
fixed point). The dynamics of the ODE for the same
setting is indicated by the dashed (green) line.
Parameter values are: $\gamma=0,\ E = 0.03, \ V_0 = 0.1,\ k=0.5,\ \mu=1$.}
  \label{gamma_0_E_003_V0_01_k_05_mu_1_min}
\end{figure}

It is well-known that in the case of the periodic (at infinity)
linearization operator that emerges in the presence of solely an
optical lattice potential, the spectrum acquires bands separated
by gaps. On the other hand, the linear potential due to its unbounded
nature at $\pm \infty$ renders the linearization spectrum of eigenstates
again a purely point spectrum (as in the more customary case of
a parabolic trap). The above discussion justifies the pure point
spectrum nature of the excitation frequencies observed
in Figs.~\ref{gamma_0_E_0004_V0_005_k_01_mu_1_min} and \ref{gamma_0_E_0004_V0_005_k_01_mu_1_max}.
However, as is well-known variations of the parameters of the system (e.g., as
was seen above, the increase of $V_0$ in the stable center case of the
effective potential minimum) can
have the consequence that they increase the soliton's
linearization frequency, which corresponds to its oscillation frequency inside
the shallow trap of its local well in Fig.~\ref{gamma_0_E_0004_V0_005_k_01_mu_1_min}.
As this parameter increases, there will emerge critical points
where the anomalous mode (of negative energy) associated with the dark
soliton will collide with the positive energy modes of the background.
These collisional events in the spectrum will produce
instabilities~\cite{review,gt07} of the oscillatory type associated with
complex eigenvalue quartets (when $\gamma=0$). An example of this
type is observed in Fig.~\ref{gamma_0_E_003_V0_01_k_05_mu_1_min}.
This is a feature of the {\it full} DGPE problem that our reduction
to a one degree of freedom ODE cannot capture. Hence, we highlight this
potential deficiency of the theory, which perceives the soliton
as a stable entity, while the dynamics of the
bottom panel of the
figure clearly illustrate the coherent structure to be oscillatorily
unstable and hence eventually departing from its local well.

\begin{figure}[tbp]
  \begin{center}
    \includegraphics[width=0.23\textwidth]{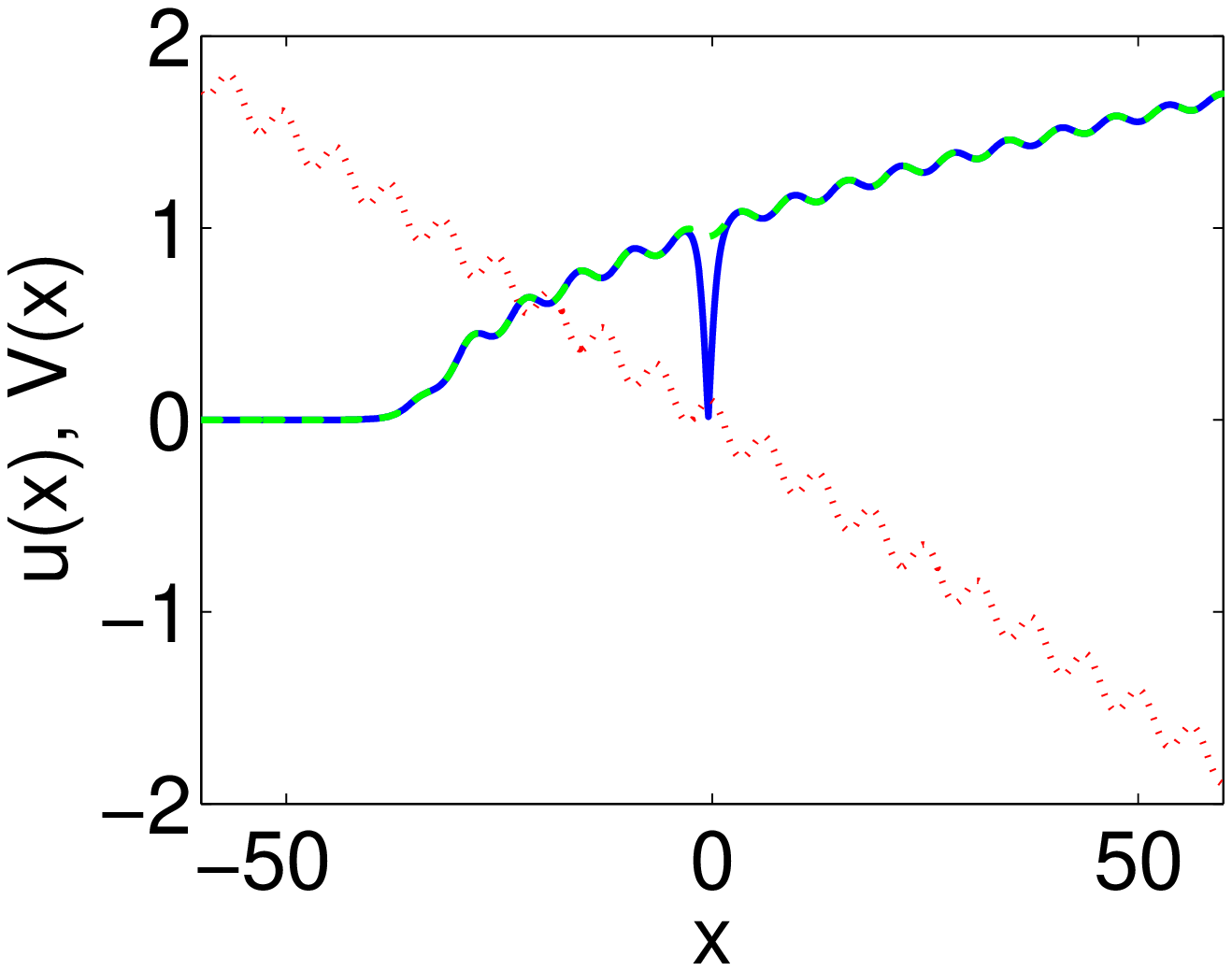}
    \includegraphics[width=0.23\textwidth]{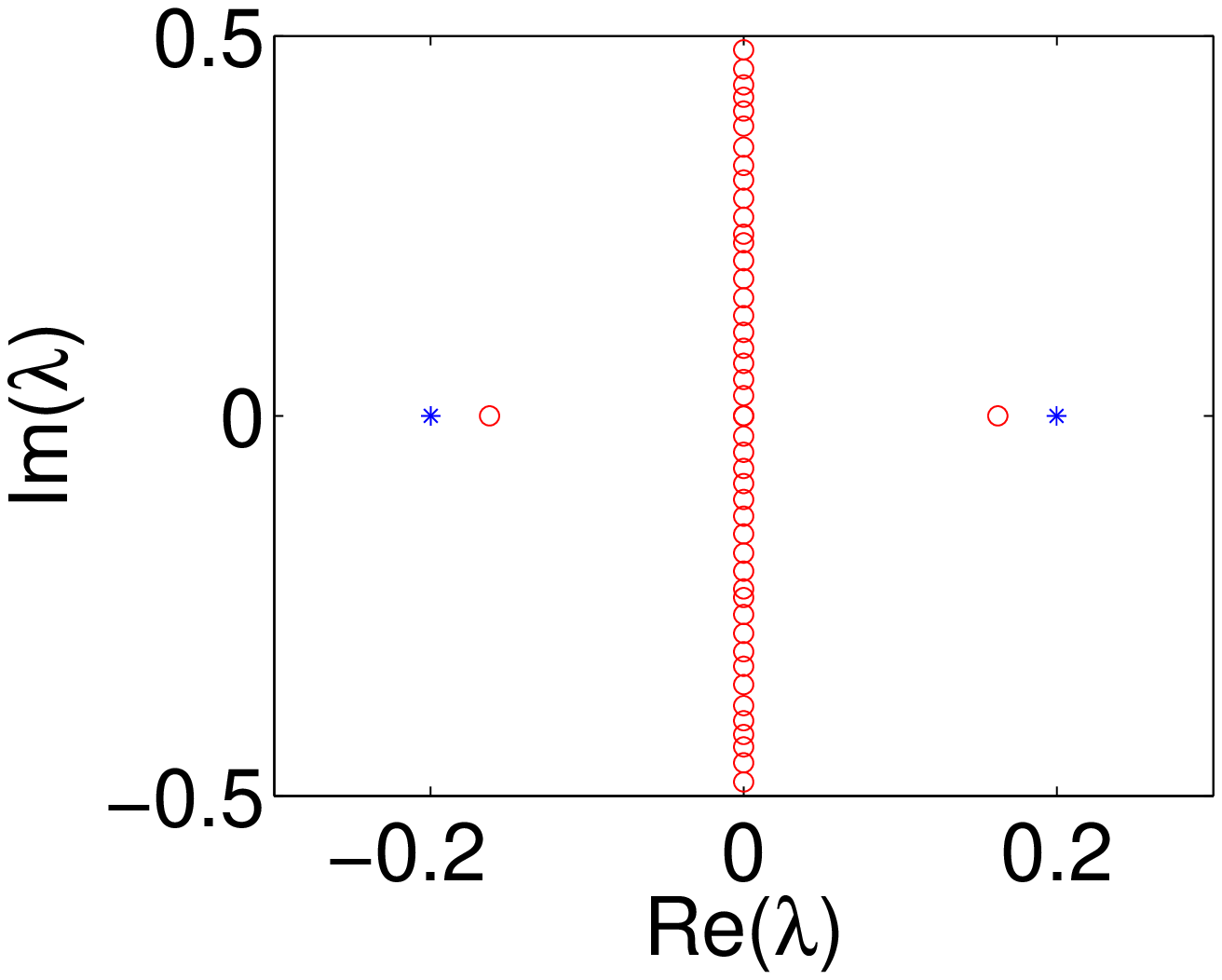} \\
    \includegraphics[width=0.33\textwidth]{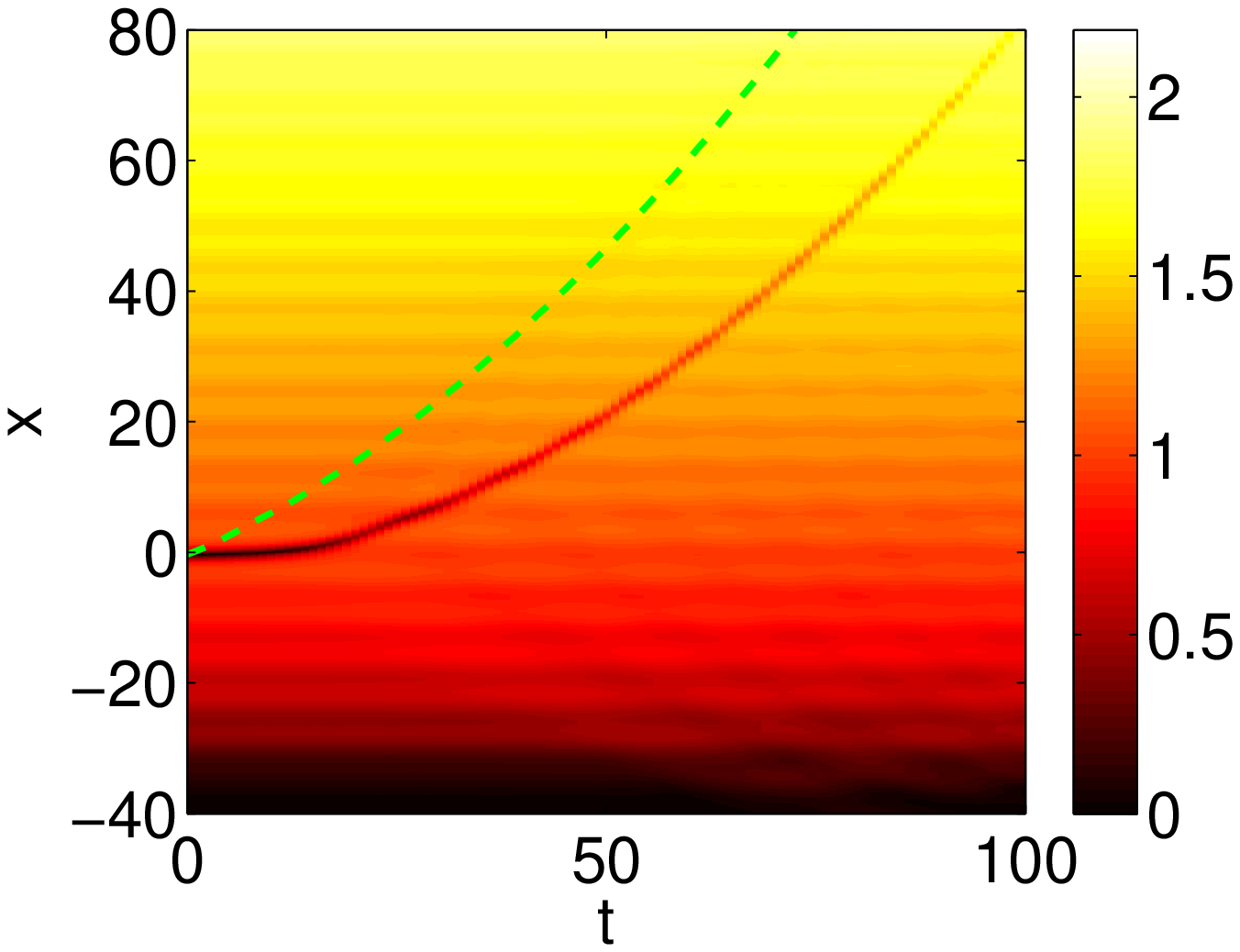}
  \end{center}\caption{(Color online) The top left panel represents the stationary
  dark soliton located at local maximum of effective external potential.
The
top right
panel shows the corresponding eigenvalues of the linear stability problem via (red) circles,
while the (blue) star represents the less successful in this case analytical prediction. Finally, the
bottom panel shows the dynamics of Eq.~(\ref{GP}) with a perturbation of the unstable saddle
point through its eigendirection of amplitude of 0.1; The dynamics of the
ODE is  indicated by the  dashed (green) line.
Parameter values are: $\gamma=0,\ E = 0.03, \ V_0 = 0.1,\ k=0.5,\ \mu=1$.}
  \label{gamma_0_E_003_V0_01_k_05_mu_1_max}
\end{figure}

Another parameter variation that renders the perturbation theory less
accurate concerns the increase of $k$.
This feature is illustrated in the
top right panel of
Fig.~\ref{gamma_0_E_003_V0_01_k_05_mu_1_max}, for $k=0.5$ and
the dark soliton being located at the local maximum of the external potential.
The difference between the unstable eigenvalue stemming from the linearization
of the PDE and that from the ODE is obviously nontrivial, and is also
clearly mirrored on the
bottom panel representing the dynamics of the
ODE and the DGPE. The over-estimate (by the ODE) of the growth rate of
the unstable eigenmode leads the dark soliton ``particle'' to diverge from
the unstable equilibrium point much faster in the particle approach
than in the full PDE one. This can presumably be attributed to the
emerging competition between the length scale of the lattice and that of
the
soliton proper (see discussion in Ref.~\cite{Theocharis}),
which contributes towards flaws
of the dark soliton perturbation theory for moderate values of $k$.

From now on, we will restrict our considerations to the regime
of small $k$ (in particular, fixing $k=0.1$),
so as to ensure the validity of the perturbation theory and avoid the
competition of the soliton and lattice length scales.


\subsection{Finite-temperature-induced dynamics}

We will now turn to a detailed examination of the role the
temperature-dependent parameter $\gamma$. A direct observation
concerning our dynamical system in the latter case
is that the Hamiltonian $\mathcal{H}(x,y) \equiv \frac{1}{2}y^2+U(x)$,
instead of being conserved as in the
case, $\gamma=0$, rather satisfies
\begin{equation}\frac{{\rm d}\mathcal{H}}{{\rm d}t} =
\frac{{\rm d}y}{{\rm d}t}y+\frac{{\rm d} U}{{\rm d}x}y=\frac{2}{3}\gamma\mu y^2 >0.
\end{equation}
Thus, the phase portrait instead of consisting of the level sets of a
constant energy $U$ (as in the case of $\gamma=0$), will rather consist
of curves that tend to higher levels of $U$ during the time evolution.

\subsubsection{Solitons located at the local minimum}

First, we consider the case of a stationary dark soliton located at the minimum
of the effective potential. As indicated above (sec.~II.B), with an increase of
$\gamma$, the stationary solutions of the
ODE and the DGPE remain unchanged, yet the linear stability properties
of such a solution
change upon crossing the critical value $\gamma_{\rm cr}$ [cf. Eq.~(\ref{gcr})]:
the local minimum of the effective potential becomes, from oscillatorily
unstable
(for $\gamma<\gamma_{\rm cr}$), exponentially unstable (for $\gamma > \gamma_{\rm cr}$);
%
these changes of stability are mirrored
in the phase plane portraits of Fig.~\ref{pp}.
%

\begin{figure}[tbp]
  \begin{center}
    \includegraphics[width=0.23\textwidth]{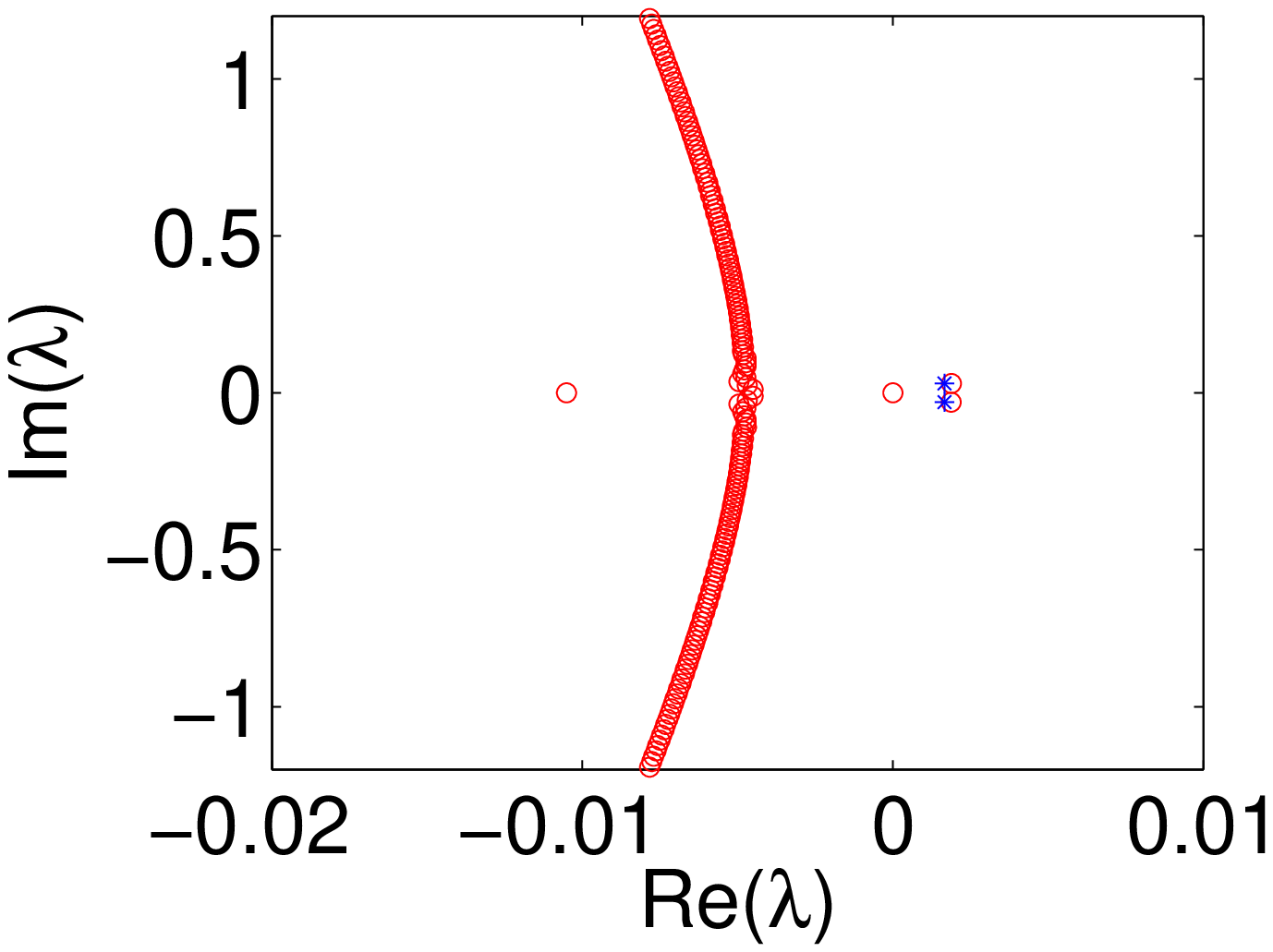}
    \includegraphics[width=0.23\textwidth]{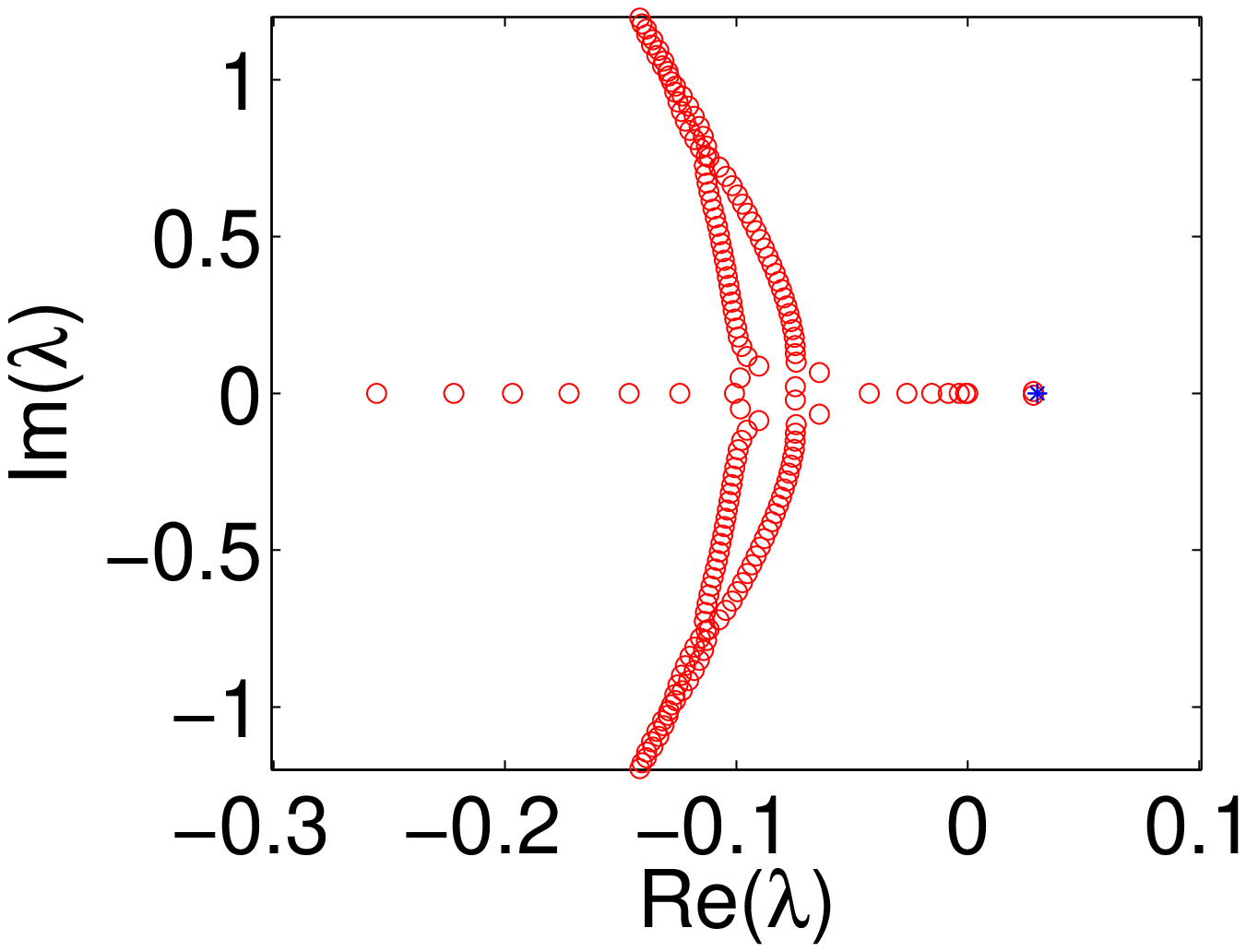} \\
    \includegraphics[width=0.23\textwidth]{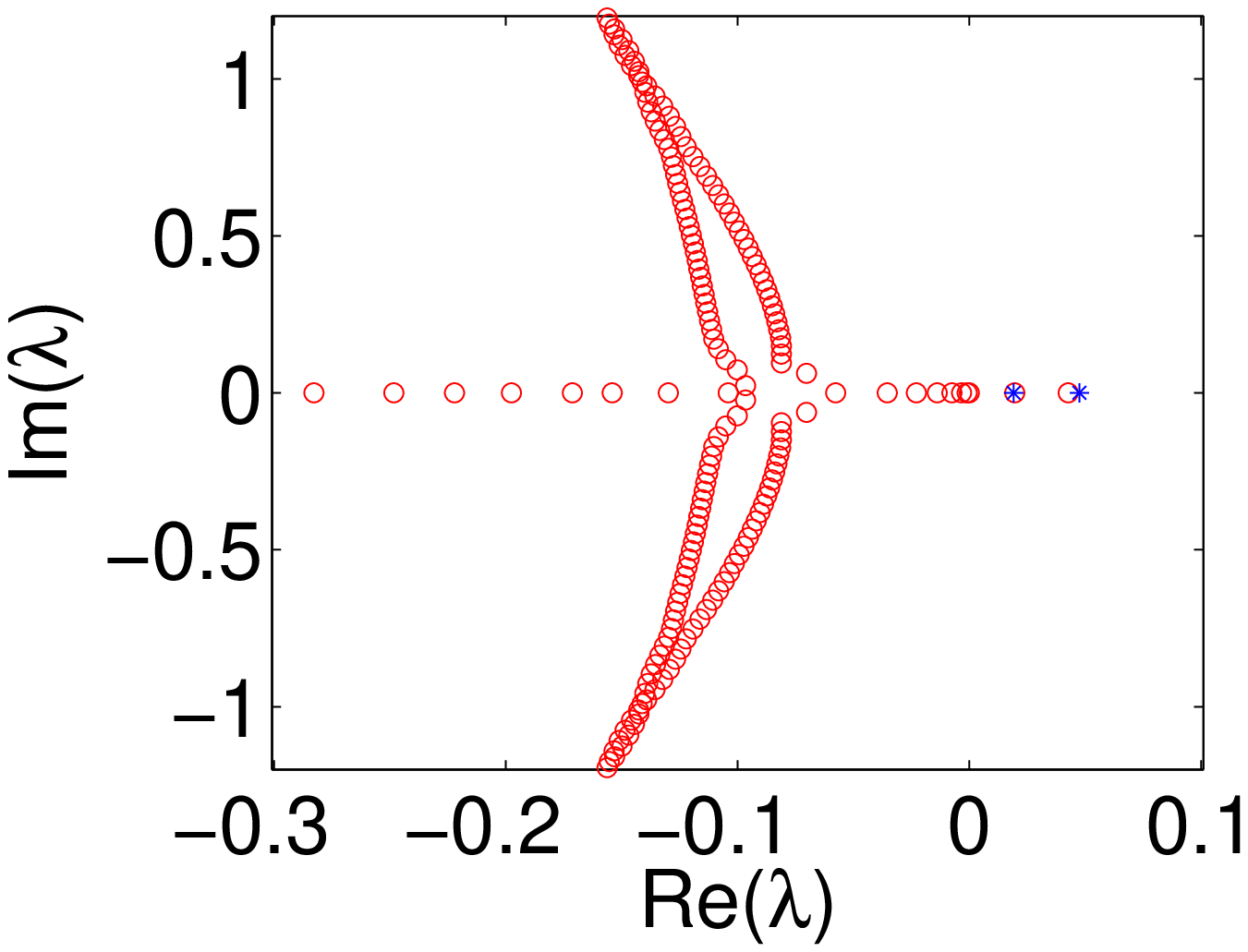}
  \end{center}\caption{(Color online) The case of the dark soliton at local minimum of the effective potential.
Three case examples of the spectral plane of the imaginary versus real part of the relevant eigenvalue
$\lambda={\rm Re}(\lambda) + i {\rm Im}(\lambda)$ is shown.
The top left panel is for $\gamma=0.005 <\gamma_{\rm cr}$, the
top right panel is
for $\gamma=0.0905=\gamma_{\rm cr}$, and the
bottom right panel is for $\gamma=0.1>\gamma_{\rm cr}$;
(blue) stars depict the analytically derived eigenvalues of Eq.~(\ref{eig2}).
Other parameter values are: $E = 0.004, \ V_0 = 0.05,\ k=0.1,\ \mu=1$.}
  \label{E_0004_V0_005_k_05_mu_1_min_eig}
\end{figure}

First, employing the BdG analysis, we will investigate if the above ODE-based
stability picture complies with the stability of the (formerly) anomalous
mode of the soliton placed at the local minimum of the effective potential.
In our simulations, we use the parameter values
$E = 0.004, \ V_0 = 0.05,\ k=0.1,\ \mu=1$, resulting in the value
$\gamma_{\rm cr}=0.0905$.
For these values, in Fig.~\ref{E_0004_V0_005_k_05_mu_1_min_eig} we illustrate
the spectral planes obtained by the BdG analysis, for three different values of $\gamma$
(below, at, and above $\gamma_{\rm cr}$).
%
It is clear that, due to the dissipative type of dynamics of the DGPE,
all eigenvalues of the operator $L$ tend to the left half
of the complex plane (thus having a negative real part and corresponding
to decaying excitations) except for the anomalous mode. The latter, as
justified in Ref.~\cite{physd}, has the opposite behavior and tends to
right half of the complex plane illustrating the unstable dynamics
of the excitation. In the figure, it is observed that the analytically
predicted relevant eigenvalues of Eq.~(\ref{eig2}) [(blue) stars] are in
excellent agreement with the anomalous mode eigenvalues [(red) circles] obtained
numerically via the BdG analysis.

As an additional interesting observation, we note the observable bending
of the stable modes in the left half of the complex spectral plane
in Fig.~\ref{E_0004_V0_005_k_05_mu_1_min_eig} and the formation of two
curves of eigenvalues for larger values of $\gamma$
presumably associated with the two different asymptotic states
within our linear plus periodic potential.

\begin{figure}[tbp]
  \begin{center}
    \includegraphics[width=0.33\textwidth]{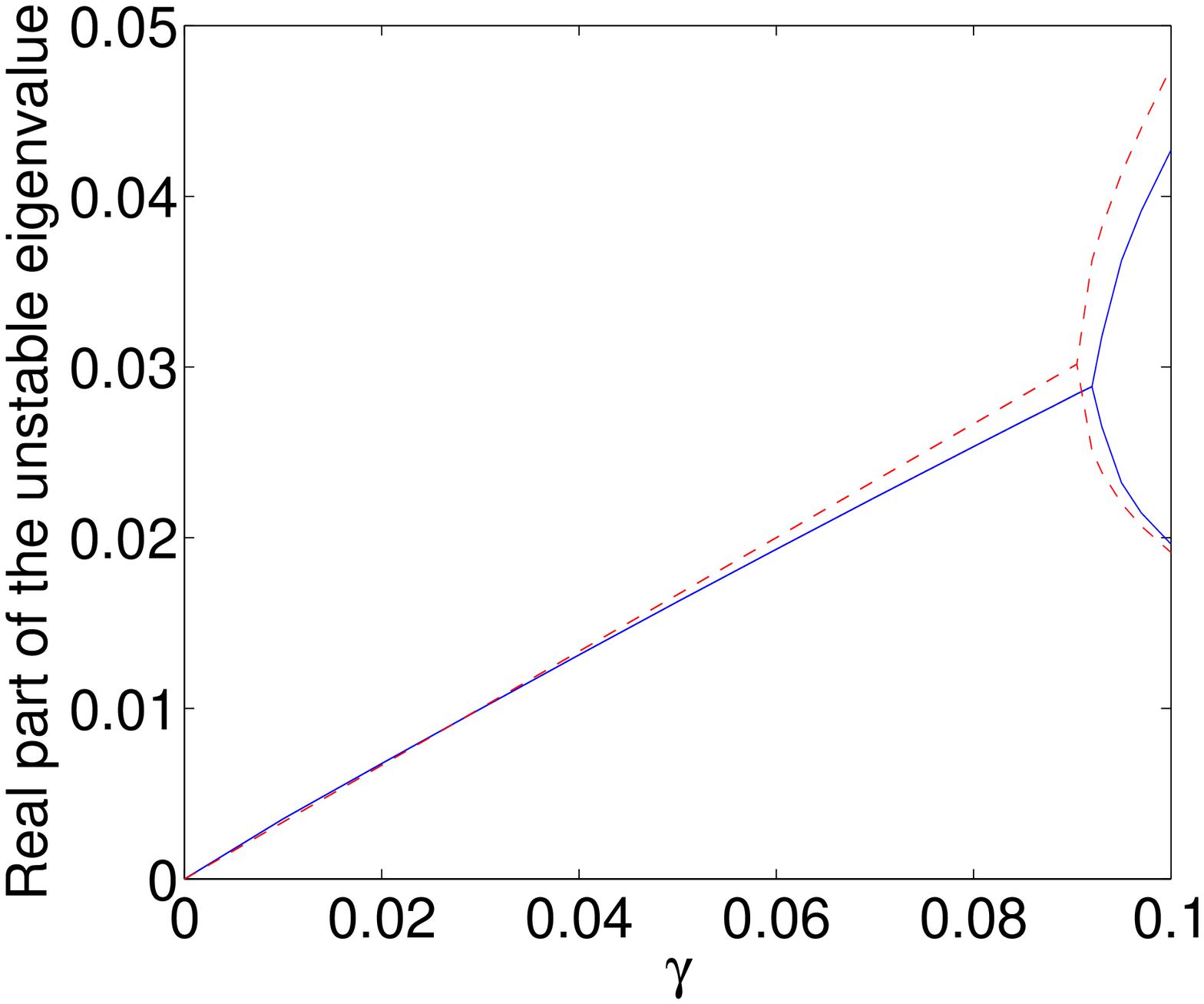}
    \includegraphics[width=0.33\textwidth]{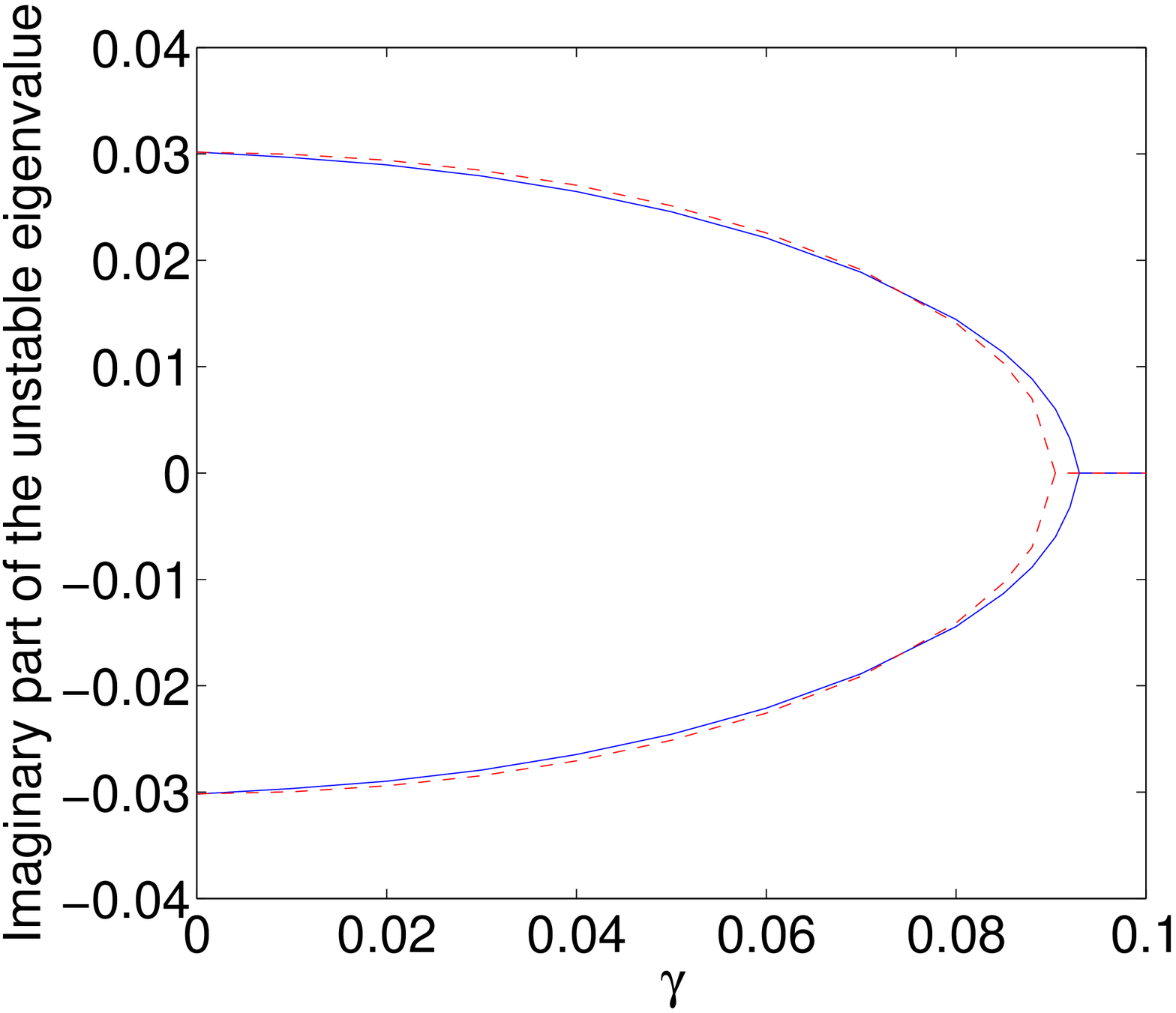}
  \end{center}\caption{(Color online)
The figure shows the real (top panel)
and imaginary (bottom panel)
part of the unstable
eigenvalue pertaining to the dark soliton located at the effective
potential minimum vs. $\gamma$. The (red) dashed line is the
result of the perturbation theory analysis,
while the (blue) solid line is the corresponding mode of the DGPE.
Parameter values are: $E = 0.004, \ V_0 = 0.05,\ k=0.1,\ \mu=1$.}
  \label{E_0004_V0_005_k_01_mu_1_eigRI}
\end{figure}

The above mentioned feature of the anomalous mode (i.e., its motion toward the right half plane) is
deeply connected to the fundamental difference between
dynamic
and thermodynamic instability of the dark soliton:
the soliton is an excited state of the
condensate and, hence, while {\it dynamically stable} for the Hamiltonian variant
of the problem, it is {\it thermodynamically unstable}. However,
in order to realize this instability and enable
the decay of the excited system to its ground state, the role of dissipation
(induced by the presence of a finite, temperature-dependent parameter $\gamma$)
is essential. This is mirrored in the direct instability of the anomalous mode,
as shown in Fig.~\ref{E_0004_V0_005_k_01_mu_1_eigRI}, in excellent agreement
with our ODE predictions for both the real and the imaginary part of
the relevant complex eigenvalue. Notice also that even the critical point for
the collision of the complex pair of eigenvalues and its subsequent
transformation into a pair of positive real
eigenvalues (when the fixed point transforms
from an unstable spiral into an unstable node) is very accurately captured.

\subsubsection{Solitons located at the local maximum}

\begin{figure}[!h]
  \begin{center}
    \includegraphics[width=0.23\textwidth]{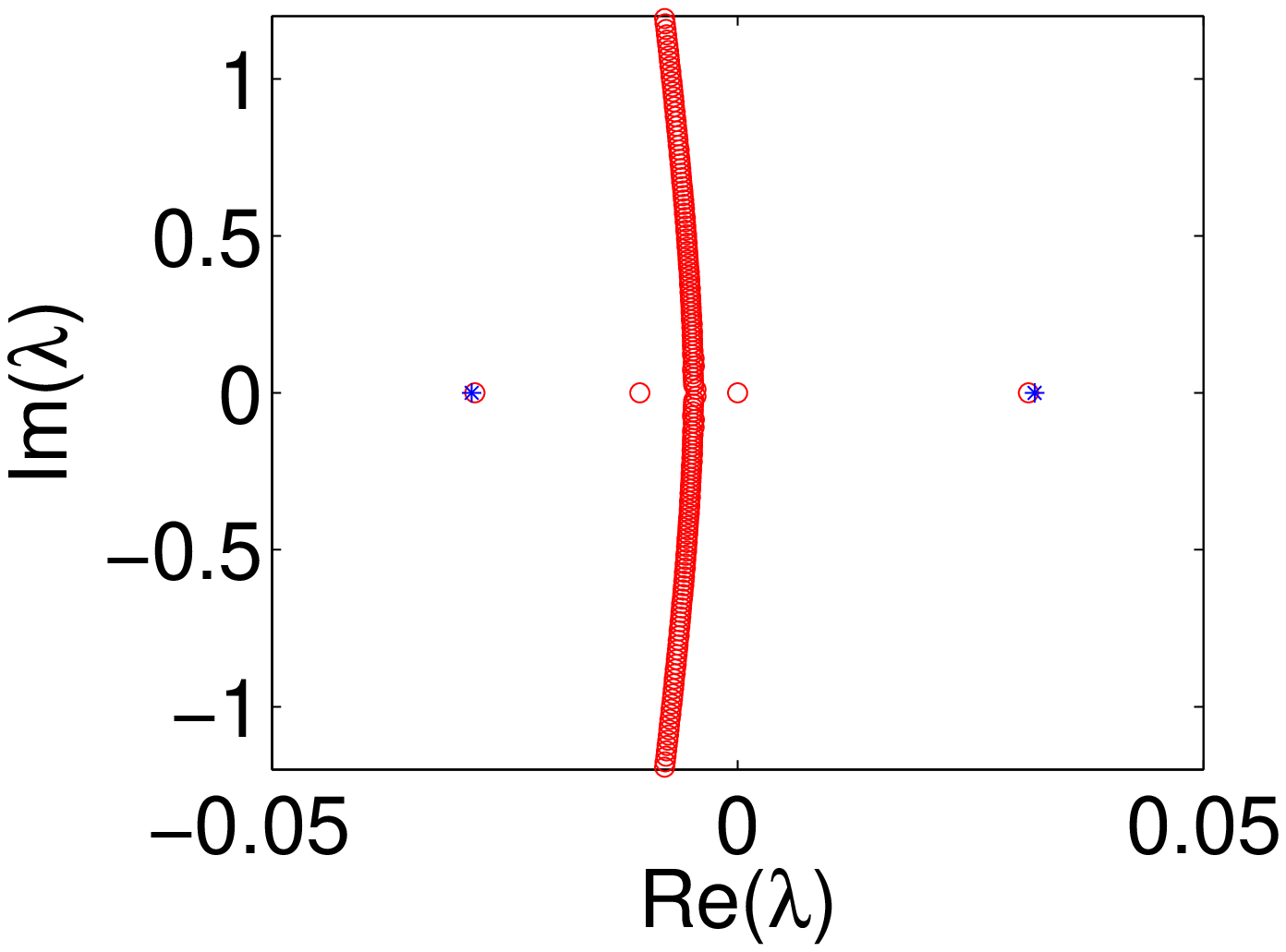}
    \includegraphics[width=0.23\textwidth]{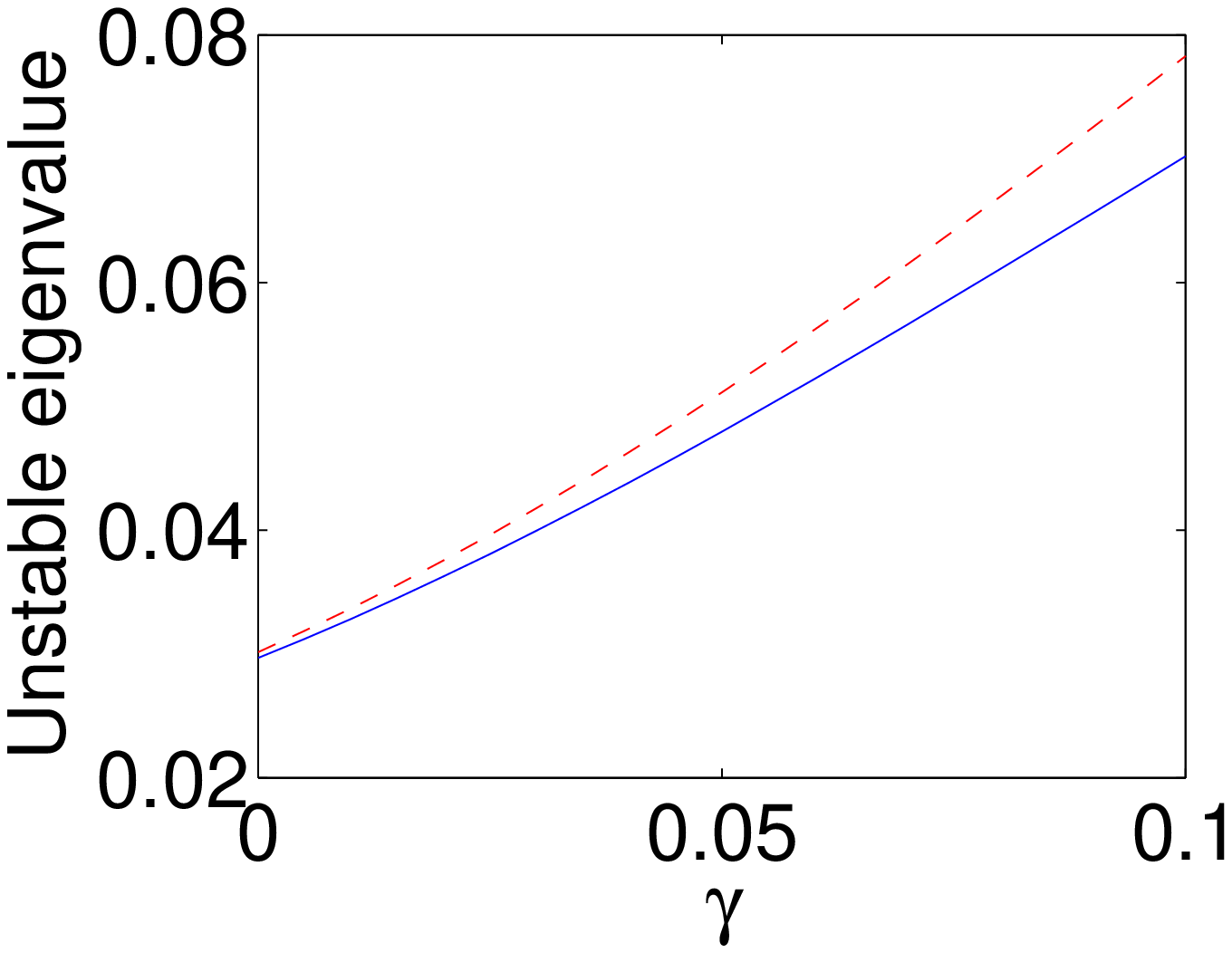}
  \end{center}\caption{(Color online) The figure shows the stability characteristics of
the dark soliton  for the case of the local max of the effective
potential. The left
panel presents a typical example of
the eigenvalues of the spectral plane
for $\gamma=0.005$. The right
panel shows the real unstable eigenvalue
of the relevant saddle point
vs. $\gamma$, with the (red) dashed line illustrating the
theoretical prediction, while the solid (blue) line is the corresponding
linear stability eigenmode of the DGPE.
Parameter values are: $E = 0.004, \ V_0 = 0.05,\ k=0.1,\ \mu=1$.}
  \label{E_0004_V0_005_k_01_mu_1_max_eig}
\end{figure}

We now turn to the examination of the effect of temperature in the case
of a soliton located at the local maximum of the effective potential.
As predicted also by the ODE, the linear stability problem for the PDE in this
case has one unstable eigenvalue (i.e., the relevant fixed
point is once again a saddle). As can be observed
in Fig.~\ref{E_0004_V0_005_k_01_mu_1_max_eig}, the agreement is once again
good for sufficiently small values of $\gamma$, but becomes less adequate
as $\gamma$ increases. Nevertheless, as can be observed, e.g., in Fig.~11
of Ref.~\cite{cock2}, physically relevant values of (the dimensionless)
parameter $\gamma$ are of $\mathcal{O}(10^{-3})$ for realistic BECs,
hence our approximation is quite reasonable in that range.
%

\begin{figure}[!h]
  \begin{center}
    \includegraphics[width=0.23\textwidth]{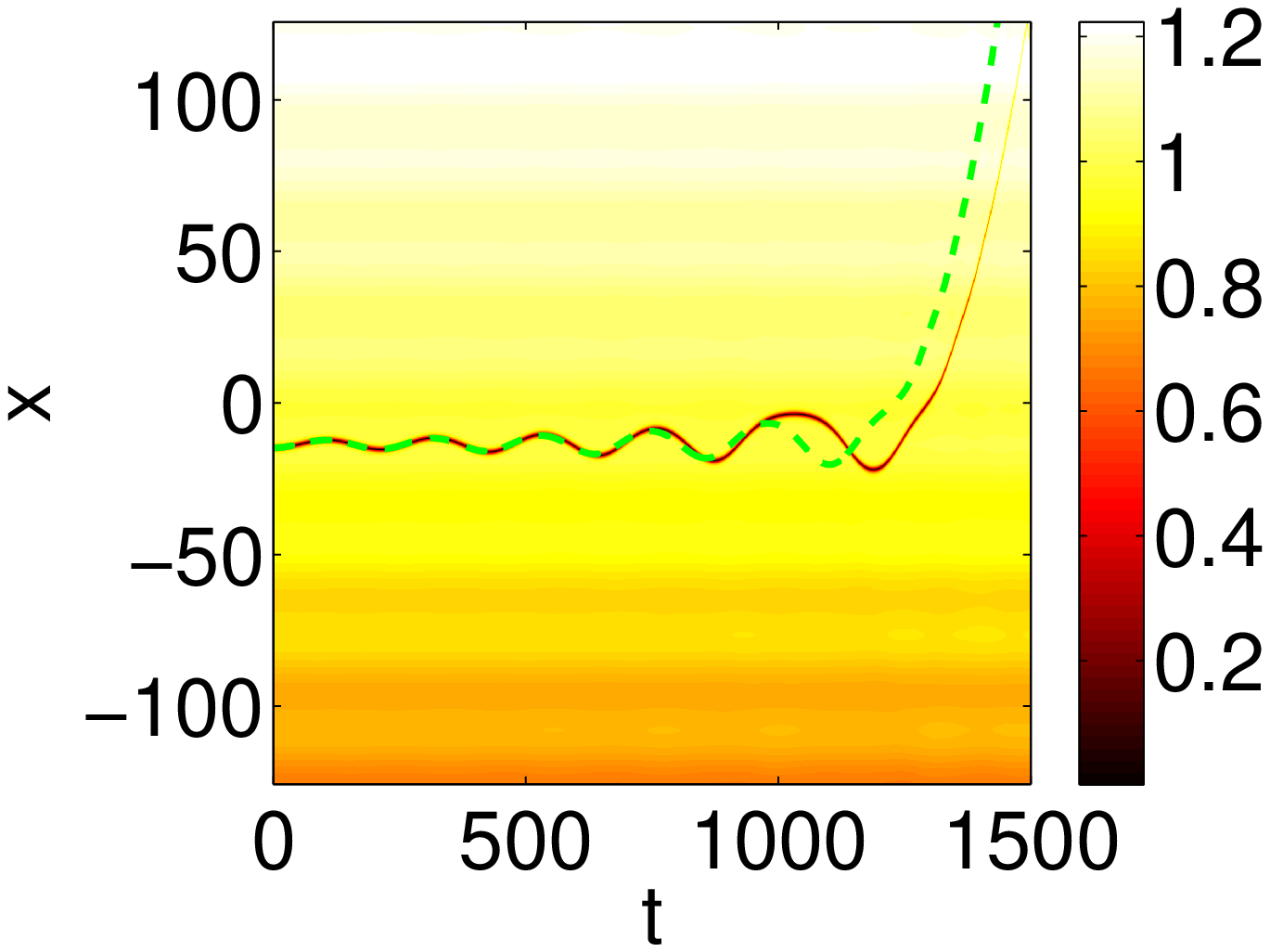}
    \includegraphics[width=0.23\textwidth]{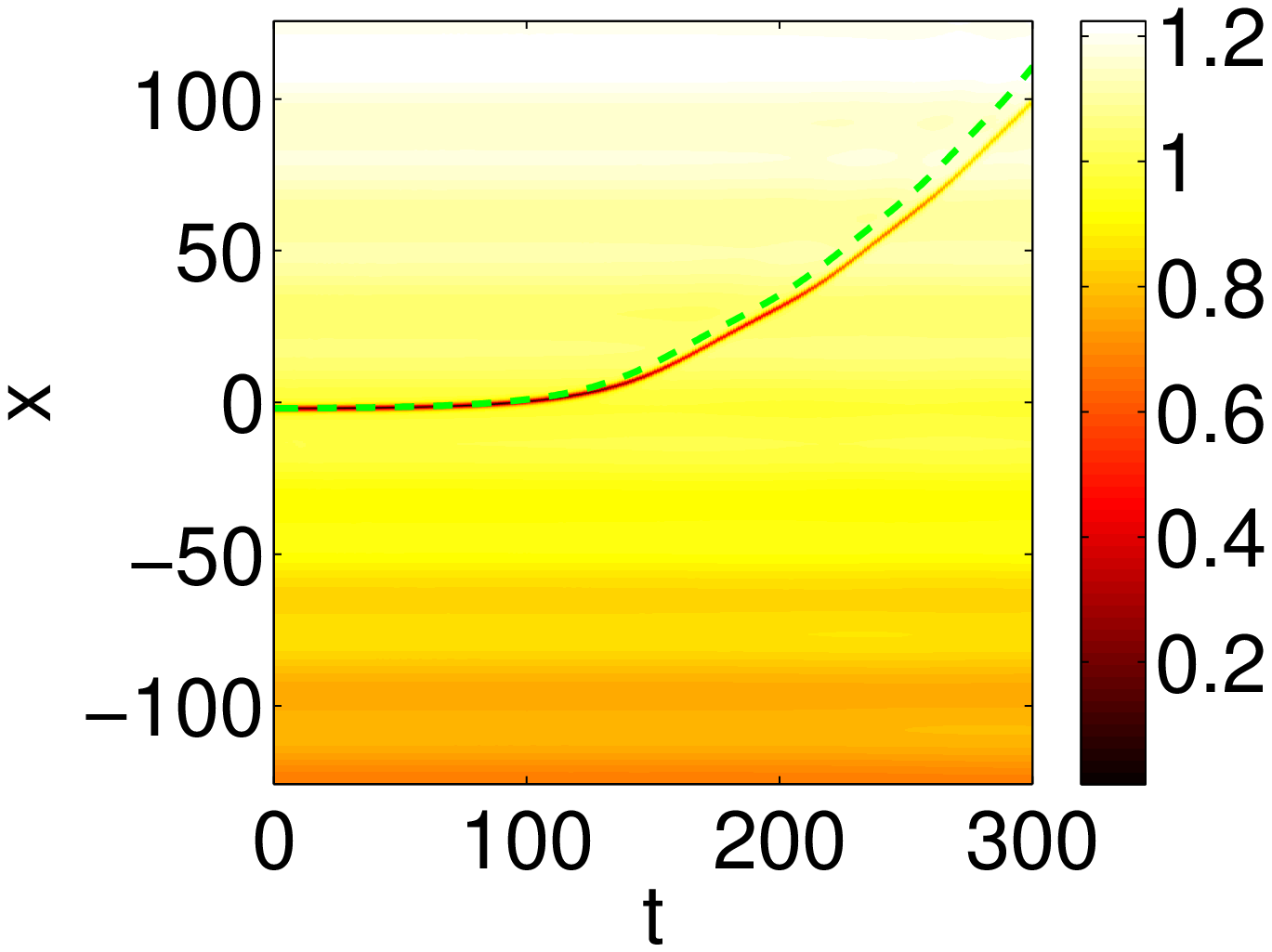}
  \end{center}\caption{(Color online) The figure presents contour plots
of the dynamics of the DGPE for parameter values
$\gamma=0.005,\ E = 0.004, \ V_0 = 0.05,\ k=0.1,\ \mu=1$.
The
left panel shows the case of the dark soliton initialized at -14.77 in the
vicinity of the effective potential energy minimum, with
the (green) dashed line showcasing the results of the perturbation theory.
The
right
panel is for the dark soliton initially located in the vicinity
of the maximum of the effective potential (at $-0.414$).
Here once again, the (green) dashed line is from the evolution of our
perturbative ODE.}
  \label{gamma_0005_E_0004_V0_005_k_01_mu_1_dynamic}
\end{figure}

Finally, as seen in Fig.~\ref{gamma_0005_E_0004_V0_005_k_01_mu_1_dynamic},
the dynamics of the unstable dark soliton in the presence of the effect of
temperature (and for values of $\gamma$ in the physically relevant range
as discussed above) is in very good agreement between the DGPE and the
corresponding ODE. This is illustrated both in the oscillatorily unstable,
for this range of $\gamma$'s, case of the minimum of the effective potential
and in the exponentially unstable case of its maximum.

\section{Conclusions}

We have studied matter-wave dark solitons in atomic Bose-Einstein
condensates with a linear and periodic (optical lattice) potential in the absence as well
as in the presence of thermal effects. Our fundamental
model employed in the analysis was the dissipative
Gross-Pitaevskii equation. Using dark soliton perturbation theory,
we found that the motion
of the dark soliton center is governed by an ODE, namely a Newtonian equation of motion,
with an anti-damping term; this ODE is strongly reminiscent of the one describing
the dynamics of the phase in a superconducting Josephson junction. We found that for
sufficiently small wavenumbers
of the periodic potential,
such that the length scale of the solitary wave
is much shorter than that of the lattice,
the ODE gives a very accurate prediction for the evolution of the center
of the dark soliton. On the other hand, for larger wave numbers
(where a competition of the scale of the
soliton and that of the periodic potential would emerge), or for
sufficiently strong potentials (beyond the realm of perturbative effects),
naturally, the approach was found to be less accurate.

In the case of Hamiltonian dynamics (zero-temperature limit), the resulting effective potential
was found to possess effective minima that were typically stable --
although potentially oscillatorily unstable -- and effective maxima,
corresponding to saddle points. The saddle points preserved their
nature under thermal perturbations, while the centers became
unstable spirals and, finally -- for sufficiently large strength
of relevant perturbation -- unstable nodes. These features
(with the exception of the oscillatory instability of the Hamiltonian
case that requires a higher-dimensional description to be captured)
were all adequately represented by our equation of motion.
In addition to the statics and near-equilibrium linearization, the ODE
was found to yield a very good approximation of the full PDE
in the corresponding  dynamics.

There are
many directions that one can envision as potential themes
for future study. On the one hand, a detailed examination of the ground
state of the system and especially of its linearization spectrum would
be a first step towards considering the ``interaction'' of this spectrum
with the anomalous mode of the solitary wave. To address such an interaction,
an expanded higher number of mode ansatz is necessary in order to capture
the relevant resonance effects. Another direction that can be envisioned
concerns a consideration of dynamical systems involving two or more solitons,
in the spirit of Ref.~\cite{andreas} in the absence,
and~\cite{us2} in the presence of thermal effects. For more such
solitary waves, more complex phenomena including the possibility
of chaotic orbits may become manifest. Another generalization may
concern the study of two-dimensional such settings, where the
study of vortices and their evolution would be of interest.
These topics will be presented in future works.

{\bf Acknowledgments.} P.G.K. gratefully acknowledges the support of the National
Science Foundation under grants DMS-0806762 and CMMI-1000337, as well
as the Alexander von Humboldt Foundation, the Alexander S. Onassis
Public Benefit Foundation and the Binational Science Foundation.
The work of D.J.F. was partially supported by the
Special Account for Research Grants of the University of Athens.



\begin{thebibliography}{99}
\bibitem{note} Although optical lattices have been argued to have notable quantum fluctuation
effects on the dynamics of dark solitons (see e.g.,~\cite{carr}) for sufficiently strong lattices
and small enough atom numbers, here we will restrict our consideration
to regimes where such effects are negligible by suitable tuning of
the OL strength (in the so-called superfluid regime)
and atomic populations.
\end{thebibliography}
\end{document}